\documentclass[lettersize,journal]{IEEEtran}
\usepackage{amsmath,amsfonts}
\usepackage{algorithmic}
\usepackage{algorithm}
\usepackage{array}
\usepackage[caption=false,font=normalsize,labelfont=sf,textfont=sf]{subfig}
\usepackage[utf8]{inputenc}
\usepackage[affil-it]{authblk} 
\usepackage{float}
\usepackage{mathtools}
\usepackage{dblfloatfix}
\usepackage{tikz}
\usepackage{amsmath}

\DeclarePairedDelimiter\floor{\lfloor}{\rfloor}
\usepackage{lipsum,booktabs}
\usepackage{amssymb}
\usepackage{graphicx}
\usepackage{comment} \usepackage{multicol}
\usepackage[labelfont=bf]{caption}
\usepackage{rotating}
\usepackage{dblfnote}
\usepackage{adjustbox}
\usepackage{epsfig}
\usepackage{ragged2e}
\usepackage{lscape}
\def\checkmark{\tikz\fill[scale=0.4](0,.35) -- (.25,0) -- (1,.7) -- (.25,.15) -- cycle;} 
\usepackage{titlesec}
\titleformat*{\section}{\large\bfseries}
\titleformat*{\subsection}{\normalfont\itshape}
\usepackage{amsmath}
\usepackage{enumitem}
\usepackage{multirow}

\usepackage{authblk}
\begin{document}
\title{Voice Spoofing Countermeasures: Taxonomy, State-of-the-art, experimental analysis of generalizability, open challenges, and the way forward}
\author[1]{Awais Khan}
\author[1]{Khalid Mahmood Malik}
\author[1]{James Ryan}
\author[1]{Mikul Saravanan}
\affil[1]{Department of Computer Science and Engineering, Oakland University, Rochester, MI 48309, USA}
\markboth{Journal of \LaTeX\ Class Files,~Vol.~XX, No.~X, August~2022}%
{Shell \MakeLowercase{\textit{et al.}}: A Sample Article Using IEEEtran.cls for IEEE Journals}
\date{}     

\maketitle
\begin{abstract}
Malicious actors may seek to use different voice spoofing attacks to fool ASV systems and even use them for spreading misinformation. Various countermeasures have been proposed to detect these spoofing attacks. Due to the extensive work done on spoofing detection in automated speaker verification (ASV) systems in the last 6-7 years, there is a need to classify the research and perform qualitative and quantitative comparisons on state-of-the-art countermeasures. Additionally, no existing survey paper has reviewed integrated solutions to voice spoofing evaluation and speaker verification, adversarial/anti-forensics attacks on spoofing countermeasures and ASV itself, or unified solutions to detect multiple attacks using a single model. Further, no work has been done to provide an apples-to-apples comparison of published countermeasures in order to assess their generalizability by evaluating them across corpora. In this work, we conduct a review of the literature on spoofing detection using hand-crafted features, deep learning, end-to-end, and universal spoofing countermeasure solutions to detect speech synthesis (SS), voice conversion (VC), and replay attacks. Additionally, we also review integrated solutions to voice spoofing evaluation and speaker verification, adversarial and anti-forensics attacks on voice countermeasures, and ASV. The limitations and challenges of the existing spoofing countermeasures are also presented. We report the performance of these countermeasures on several datasets and evaluate them across corpora. For the experiments, we employ the ASVspoof2019 and VSDC datasets along with GMM, SVM, CNN, and CNN-GRU classifiers.
(For reproduce-ability of the results, the code of the testbed can be found at our GitHub Repository\footnote{{https://github.com/smileslab/Comparative-Analysis-Voice-Spoofing}}).

   \end{abstract}
   
   
\section{Introduction}
\IEEEPARstart{A}{utomatic} speaker verification systems (ASVs) are now in wide use for authentication, such as for over-the-phone banking, and are becoming an increasingly important biometric solution in the current climate of the pandemic to limit the spread of disease. Although ASVs can be used to validate an identity, they are prone to spoofing attacks, which may allow an unauthorized user to gain access to privileged information such as a bank account. Some of these attacks, such as speech synthesis and voice conversion, are also being used in cyberspace to spread misinformation and disinformation. In addition, ASV systems are more prevalent in IoT devices, such as smart speakers. A spoofing attack on IoT devices connected in a chain to control someone's home or other devices operated via the ASV system has the potential to allow unauthorized access to security devices, e.g., IoT door locks \cite{malik2020light}. 
\par Spoofing attacks on ASVs can be grouped into Physical Access (PA) attacks and Logical Access (LA) attacks \cite{balamurali2019}. An example of a PA attack is a replay attack, where the hacker records the original speaker's voice using any recording device without consent. Later, this prerecorded voice can be replayed onto the ASV system in order to compromise the voice bio-metric based access control in financial institutions and smart homes \cite{malik2020light}. LA attacks, conversely, are comprised of artificial, or machine-generated, cloned samples. LA attacks consist of text-to-speech (TTS) synthesis and voice conversion (VC). For TTS synthesis, cloned samples are generated using original speech and transcripts, while VC attacks use only voice samples of the original speaker to train different deep learning models, i.e., a vocoder, waveform, etc., to generate the cloned samples. Again, the objective is to generate realistic voice samples of a target speaker in order to compromise the security of an ASV system, and thus gain access to someone's home, bank account, or other voice-controlled application.
Due to a massive escalation in the availability of high-quality recording equipment and the ease of their creation, replay attacks are being generated more easily and rapidly, even by less tech-savvy people, and can be classified into single-hop or multi-hop attacks \cite{baumann2021voice}. More specifically, a single-hop replay attack utilizes a single recording device to play the audio sample of a verified user. In contrast, a multi-hop replay attack utilizes multiple recording devices (i.e., a verified user's voice is recorded, then replayed to another recording device, and then replayed to the ASV system) in a chained scenario. In our prior work \cite{javed2021towards}, we reported a new hybrid voice spoofing attack, i.e., a cloned replay, where the voice of a target speaker is captured and replayed to an ASV system, that can also be generated to spoof an ASV system. A more recent threat to ASVs and spoofing counter measures is adversarial machine learning \cite{szegedy2014intriguing}, where machine learning algorithms may be deceived by injecting minor distortions into the audio sample. According to Szegedy et al. \cite{szegedy2014intriguing}, deep learning model predictions can be easily modified by extremely small perturbations.
 
Several countermeasures have been proposed to defeat voice spoofing attacks, and these are often comprised of two parts: the first one (front end) is the feature representation scheme for the input speech signal, and the second one (back end) is a classifier to distinguish between bonafide and spoofed samples. The feature descriptor (front end) should be capable of effectively capturing the traits of the dynamic vocal tracts of a bonafide speaker. Similarly, the back-end classifier should be able to better learn the distinct traits of bonafide and spoofed speech samples in order to accurately discriminate against the spoofed speech samples. In contrast to the traditional approach of front-end and back-end solutions, in the past few years, the research community has focused on deep learning and end-to-end solutions to combat voice spoofing attacks. Existing countermeasures have only tackled single-hop spoofing attacks, and chained-replay attacks and their countermeasures have been largely overlooked \cite{9107388}. Recent efforts toward in finding a unified solution to the problem of speaker verification, where a single countermeasure may be applied to several attacks, have been studied. Although these countermeasure solutions are only beginning to be explored, there is a marked need for a unified solution as the way forward in ASV anti-spoofing techniques \cite{zhang2022probabilistic}  \cite{jung2022sasv}.
\begin{table*}[!b]
\centering
\caption{Comparison of the existing survey and review papers}
\label{tab:survey}

\begin{tabular}{*{10}{c}} 
\hline
 Paper & \multicolumn{3}{l}{Presentation Attacks} & \multicolumn{3}{l}{Countermeasures} & Integrated ASV & Experimental Analysis & Cross Corpus\\
 \hline  \hline
  & Replay & VC/Synthesis & Adversarial & Hand Crafted & Deep Learned & End-to-End & & &
 \\ \hline
 \cite{wu2015spoofing} & \checkmark & \checkmark & $\times$ & \checkmark & \checkmark & \checkmark & $\times$  & $\times$  & $\times$ \\  \hline
 \cite{sahidullah2016integrated}  & \checkmark & $\times$ & $\times$ & \checkmark & $\times$ & $\times$ & $\times$  & \checkmark & $\times$ \\ \hline
 \cite{font2017experimental} & \checkmark & $\times$& $\times$ & \checkmark & $\times$ & $\times$ & $\times$  &  \checkmark &  \checkmark\\ \hline
  \cite{patil2018survey} & \checkmark &  $\times$ & $\times$ & \checkmark &  $\times$ &  $\times$ & $\times$  & $\times$  & $\times$ \\  \hline
\cite{sahidullah2019introduction} & \checkmark & \checkmark & $\times$ & \checkmark &\checkmark  & \checkmark & $\times$  & $\times$&  $\times$\\ \hline
\cite{tan2021survey} & \checkmark& \checkmark&$\times$&\checkmark & \checkmark&\checkmark & $\times$  & $\times$&  $\times$\\ \hline
\cite{mittal2021automatic} & \checkmark& \checkmark & $\times$ & \checkmark&\checkmark &\checkmark & $\times$  & $\times$&  $\times$\\ \hline
\cite{kamble2020advances} & \checkmark & \checkmark & $\times$ & \checkmark & \checkmark  & \checkmark & $\times$  & $\times$ &  $\times$\\ \hline

 [Ours] & \checkmark & \checkmark & \checkmark & \checkmark & \checkmark &\checkmark  & \checkmark & \checkmark & \checkmark \\ \hline
    \end{tabular}
\end{table*}
The surveys on ASV systems and spoof detection techniques conducted to date have primarily been focused on specific spoofing attacks and the countermeasures for them. To the best of our knowledge, six articles \cite{wu2015spoofing, patil2018survey, sahidullah2019introduction, kamble2020advances, tan2021survey, mittal2021automatic} have been presented as surveys, and two studies \cite{sahidullah2016integrated} and \cite{font2017experimental} conducted a comparative analysis of voice spoofing countermeasures. 
\par In a survey of voice spoofing countermeasures, Wu et al. (2015), in \cite{wu2015spoofing,wu2015sas,wu2015asvspoof}, provided a good classification of voice spoofing attacks known at the time, along with ASV system vulnerabilities. However, the age of this review means the advanced spoofing attacks that have grown prevalent in ASV systems were not addressed. In addition, is focused on a dedicated countermeasure for each sort of attack, rather than exploring a more unified solution. Following this, Sahidullah et al., \cite{sahidullah2016integrated} and Font et al., \cite{font2017experimental} published comparative analyses of the spoofing countermeasures that focused solely on replay attacks. According to their work, published during 2017–2018, replay attacks attracted the most attention in the wake of the ASVspoof 2017 challenge. The results of \cite{sahidullah2016integrated}, using the ASVspoof2015 dataset, demonstrated that an ensemble of acoustic features, together with machine learning models, produced more accurate results than individual classifiers.  Kamble et al. \cite{kamble2020advances} also published a work that covered a subset of the specific speech corpora along with the evaluation measures in this field. 
\par Font et al. \cite{font2017experimental} performed a comparative analysis of nine different countermeasures with cross-corpus evaluation of the countermeasures. However, the authors only discussed the performance of the countermeasures against replay attacks due to the nature of the dataset used for the study. Although Sahidullah's (2016) \cite{sahidullah2016integrated} study and Font's (2017) \cite{font2017experimental} paper each provided a comprehensive analysis of the replay attack, other types of spoofing attacks that could severely disrupt an ASV system were ignored. Sahidullah et al. surveyed four different types of spoofing attacks, spoofing procedures, and countermeasures the next year, in 2019 \cite{sahidullah2019introduction}, summarized some spoofing challenges and presented countermeasures. However, a thorough examination of all aspects of the ASV system was still wanting. For instance, \cite{sahidullah2016integrated} and \cite{font2017experimental} focused only on front-end countermeasure design, and  concentrated on the SS, VC, replay, and mimicking attack types, alone where speech corpora, protocols, classifier, and evaluation metrics all contributed equally to the construction of an ASV system. These were absent from the \cite{sahidullah2019introduction} study. 
\par Aside from \cite{tan2021survey}, most of the articles did not present a detailed taxonomy of modern voice presentation attack detection (PAD) methods, where the work was structured based on identified elements. This survey, along with \cite{mittal2021automatic} broadened the classification of relevant work and built on the taxonomy from the most recent work on voice PAD. The significant contribution of these articles was the presentation of trends and analyses of voice PAD that were absent in the other survey articles. Mittal et al. (2021) described the computation mechanisms of traditional and modern speech feature extraction, as well as datasets and a combination of various front- and back-end techniques. None of these articles, however, performed a fair experimental analysis of the existing state-of-the-art countermeasures. In addition, existing reviews and survey articles also lack cross-corpus evaluation in order to show the generalizability of existing solutions. The comparison of the existing reviews and surveys is presented in Table \ref{tab:survey}.

\par Existing spoofing countermeasures have employed diverse features and classifiers and, typically, have had  their performance evaluated on only one of several datasets, i.e., ASVspoof 2017, ASVspoof 2019, and ASVspoof 2021, using different metrics. With no standardization, it is difficult to declare a single countermeasure as the unified method that works best. Thus, there exists a need to provide a thorough analysis of existing countermeasures in order to show which method is the best fit for a certain scenario, i.e., spoofing type, dataset, single or multi-hop attack, etc. To the best of our knowledge, no comparative analysis work on multiple voice spoofing attacks, including single and multi-order attacks, has ever been presented. Moreover, existing studies have ignored the important aspect of cross-corpora evaluation, which is crucial to the evaluation of the generalized nature of the countermeasure. To address these challenges for the existing countermeasures, we present a comparative analysis of various spoofing countermeasures. Motivated by discussions, the survey proposed in this paper investigates significant contributions to the ASV system development chain. More specifically, the main contributions of our work are:
 \begin{itemize}
  \item We present a baseline survey of voice spoofing countermeasures, involving diverse factors, for the benefit of upcoming anti-spoofing systems.
 \item
 We present a comprehensive analysis of existing state-of-the-art voice spoofing attacks, countermeasures (hand crafted features, along with deep learning and end to end solutions), publicly available datasets, and the performance evaluation parameters used in voice spoofing. 
 \item
 We present an experimental analysis of state-of-the-art different countermeasures, on several classifiers, for spoofing detection performance using the VSDC, ASVspoof 2019 and 2021 datasets, and evaluate the results across corpora.
 \item
 We cover one of the key limitations of cross corpus evaluation and generalization mentioned by the in-field researchers and evaluate the performance of the featured countermeasures on three large-scale publicly available and diverse datasets, i.e., ASVspoof 2021, 2019 and VSDC, in terms of min-tDCF and EER. The countermeasures are tested against four different machine learning and deep learning classifiers.
\end{itemize}
\begin{table*}[t]
 \caption{\label{tab:PRISMA-table}Literature collection protocol.}
\centering
\begin{tabular}{ p{4cm} | p{12cm} }
  \hline  \hline
 Preparation Protocol & Description\\
 \hline
 Purpose &  \begin{itemize}[noitemsep,nolistsep,nosep]
     \item To identify current state-of-art voice spoofing attacks and countermeasures.
     \item To critically compare both popular approaches and the features used in voice spoofing countermeasures.
     \item To investigate open challenges that still need investigation in the domain of voice spoofing countermeasures, and paths forward.
 \end{itemize} \\
 \hline
Sources &  
     Google Scholar, IEEE explore , Springer Link\\
\hline
 Query &  Queries were used on the data sources above for collection of sources: Audio Spoof Countermeasure/ Spoofed Audio Detection/ Audio Synthesis Detection/ Audio Replay Detection/ Audio Spoofing Countermeasures/ Automatic Speaker Verification/ Secure ASV System/ ASV spoofing/ Presentation Attacks/ Voice Conversion/ Adversarial Attacks on ASV/ Voice Replay Attacks/ Anti Spoofing Countermeasures/ Voice Spoofing Detection.  \\\hline
 Method &  Literature was categorized as follows: \begin{itemize}[noitemsep,nolistsep,topsep=0pt, nosep]
     \item Audio spoofing detection and countermeasures based on traditional and deep-learned methods, as well as handcrafted features.
      \item Existing voice spoofing attacks, along with adversarial attacks and their taxonomy.
     \item An extensive examination, testing and comparison of existing voice spoofing countermeasures using single and cross corpus evaluation. 
     \item Detailed discussion of the research gap, issues, limitation, and future trends in voice spoofing detection and countermeasures.
 \end{itemize}\\
 \hline
 Size & A total of 150 papers were retrieved using this query mentioned above from the method until the search was discontinued on 04-20-2022. We selected literature that was relevant to the subject of audio spoof detection and excluded those that were not relevant to this subject or were white papers/articles.  \\
 \hline
 Inclusions and Exclusions & Preference was given to peer-reviewed journal papers and conference proceedings articles. In addition, articles from the archive literature were also taken into account.\\
 \hline

 \end{tabular}
\end{table*}

\section{Literature collection and selection criteria} 
\begin{figure}[!b]
    \centering
    \includegraphics[width=8 cm]{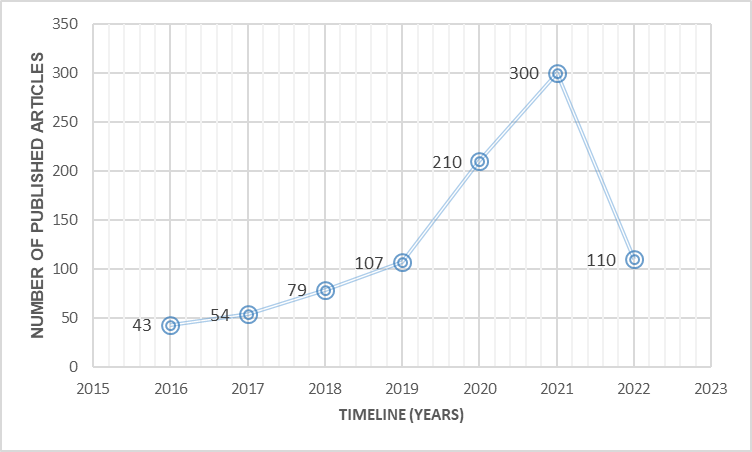}
    \caption{Number of papers in the area of voice Spoofing Attacks and Countermeasures by yearly publication.}
    \label{fig:Figure1}
\end{figure}

In this survey, we review existing research papers that focus on techniques for detecting and countering audio spoofing attacks. A detailed description of the approach and protocols employed for the review is given in Table \ref{tab:PRISMA-table}. We also display trends in spoofing attacks and countermeasures year over year by examining Google Scholar papers released in the previous 6 years (2015–2022). Figure \ref{fig:Figure1} depicts the course of the published study. The rest of the paper is structured as follows: Section 3 provides current voice spoofing countermeasures. Section 4 discusses voice spoofing attacks as well as adversarial attacks, and Sections 5 and 6 discuss publicly available data sources and performance evaluation strategies. Section 7 describes the acoustic features and classifiers used in a comparative analysis of the countermeasures. Sections 8 and 9 present the experimental setup and analysis, as well as a discussion of limitations and future trends. Finally, Section 10 provides our conclusions and motivation for future work.  

\section{Voice Spoofing Attacks}
\begin{figure*}[t]
    \centering
    \includegraphics[width=13 cm]{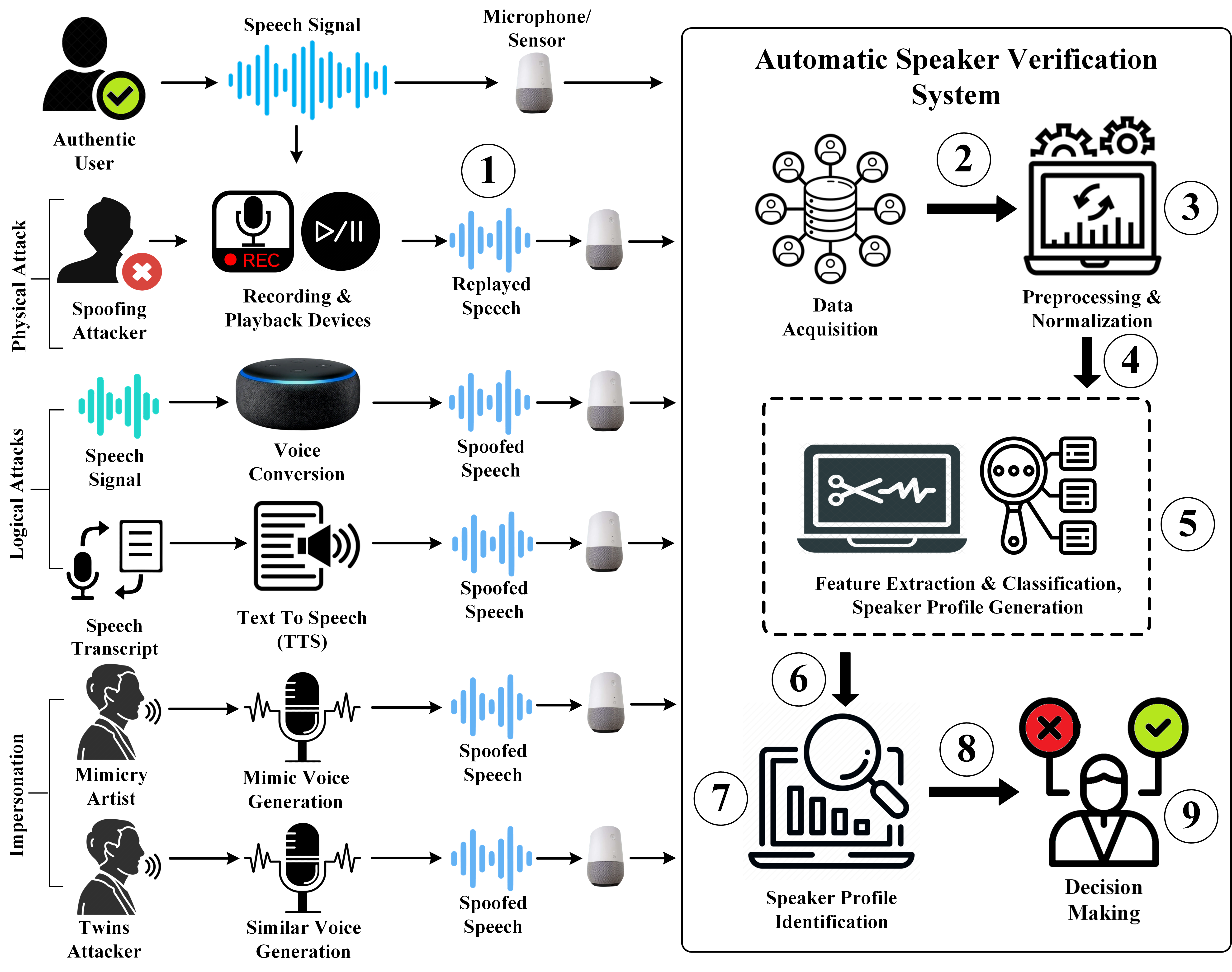}
    \caption{Existing Threats to Automatic Speaker Verification Systems.}
     \label{ASV_Final}
\end{figure*}

 ASV systems, are vulnerable to a variety of voice spoofing attacks, i.e., physical-access (PA) and logical-access (LA) attacks. These and other possible attacks on ASV systems are graphically illustrated in Figure \ref{ASV_Final}. PA attacks are easier to conduct, compared to LA attacks, due to the availability of modern high-quality recording devices. A replay attack can be generated using a device as simple and commonly available as a mobile phone, compared to the robust and sophisticated AI voice cloning algorithms required for LA attacks. Hence, voice replay attacks are more commonly employed to spoof ASV systems in order to accomplishing an attacker's objectives e.g., to gain entry to someone's home. Moreover, we have demonstrated in our earlier work \cite{baumann2021voice} that voice replays can be generated not only in a single order but also in multi-order attacks, by using different smart speakers such as Amazon Alexa and Google Home.

\begin{figure*}[t]
    \centering
    \includegraphics[width=15cm]{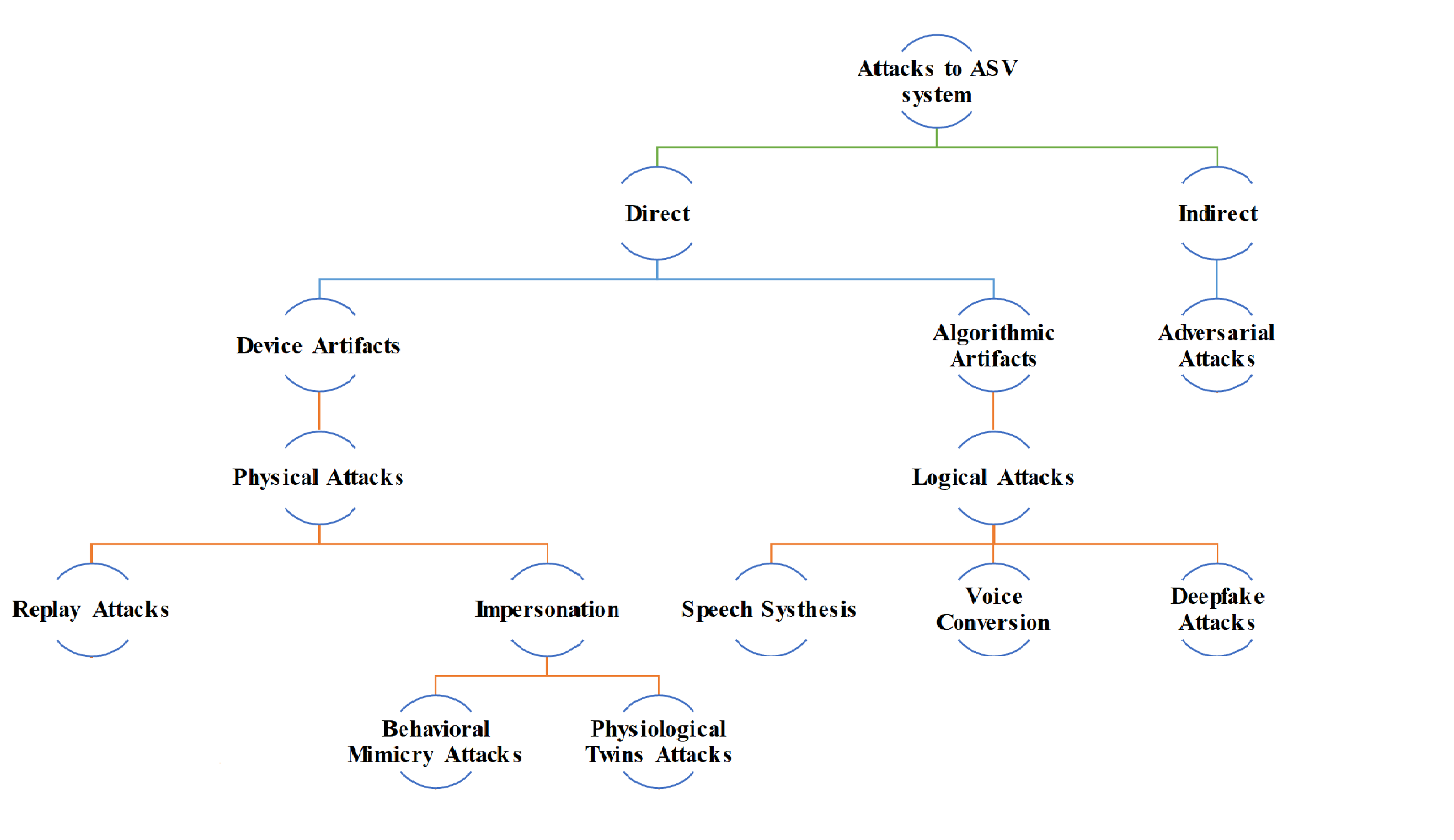}
    \caption{Taxonomy of the Voice Spoofing Attacks.}
     \label{attacks}
\end{figure*}

Voice spoofing countermeasures, also known as presentation attack detection (PAD) systems, aim to detect the following attacks (see also Figure \ref{attacks}):
\begin{itemize}
    \item Direct attacks can further be broken down into device artifacts and algorithmic artifacts, which are \textbf{A)} Physical access (PA) and \textbf{ B)} Logical Access (LA) scenarios, respectively.
    \begin{itemize}
        \item  PA attacks occur when the samples are applied as an input to the ASV system through the sensor, such as a microphone.
        \begin{itemize}
            \item Replay attacks are one example of ASV PAs. Replay attacks occur when the audio of an authorized user is recorded and played back in order to deceive an authentication system. Replay attacks are the easiest to attempt, as they do not require any special technical knowledge, and can easily deceive ASV systems. Because replayed audio includes background and/or other noise, however, it can be detected. 
            \item An impersonation attack occurs at the microphone level, where a person changes how they talk to mimic the speech characteristics of a legitimate user. An attacker can fool an ASV system if the impostor’s natural voice has similar features.
        \end{itemize}
        \item LA attacks occur when the audio samples bypass the sensor and are injected directly into the model. Spoofing attacks based on logical access include \textbf{C)} voice conversion, \textbf{D)} speech synthesis, and E) Adversarial attacks.
            \begin{itemize}
                \item Voice conversion uses an imposter’s natural voice to generate artificial speech in order to match the targeted speaker's voice. Attacks are usually created by other models to fool a specific ASV system. Machine learning has allowed the mapping of speech features between speakers to be accurate, and is now computationally efficient enough to make ASV systems vulnerable to these types of attacks. However, these attacks can still be detected because they are not a perfect match to genuine audio. 
                \item Speech synthesis attacks, also known as deepfake attacks, are similar to voice conversion but use text as an input to the model in order to generate a voice clip similar to a targeted speaker in an effort to fool an ASV system. Models that generate accurate speech features can be trained using a small data set of recorded audio.
            \end{itemize}
            
        \end{itemize}
    \item Indirect attacks include adversarial attacks, where  the audio signal remains unchanged while the attacker modifies the signal's properties during ASV processing. The recent work on adversarial attacks is discussed later in the paper. 
\end{itemize}

\subsection{Proposed countermeasures to Adversarial Attacks}
Adversarial attacks, such as the addition of Gaussian noise, can create small perturbations in audio samples and pose a threat to ASV anti-spoofing models, affecting accuracy and safety. The method described in \cite{wu2020defense} uses self-supervised learning to leverage the knowledge of unlabeled data, which significantly improves performance. They create high-level representations, extracted by the self-supervised model, and a layer-wise noise to signal ratio (LNSR) in order to quantify and measure the effectiveness of deep models in countering adversarial noise. During training, different masking methods are applied to 15\% of the audio samples to create a model that reconstructs audio. This defense is passive, i.e., not a proactive defense. Their proposed method, called Mockingjay, filters the audio. This model was only tested on the Projected Gradient Descent (PGD) and Fast Gradient Sign Method (FGSM) attacks but not on other attacks. Also, as epsilon (the strength of an attack) increases, the defense starts to weaken and eventually fails. Both white-box and black-box scenarios were tested. 
\par Another defensive method, \cite{wu2020defense2}, uses spatial smoothing, filtering, and adversarial training. This method has only been  tested against PGD attacks and may perform poorly under other, perhaps stronger, attacks because different attacks leave different patterns on the sample. These defense methods improved the robustness of spoofing countermeasures against PGD attacks, but they are still limited in their overall defensive capability because adversarial training becomes weaker against unfamiliar attacks. The work of \cite{wuCNN} creates a model that fits the distribution of genuine speech, where it takes genuine speech as the input and generates "genuine" speech, or output with the same distribution as the genuine speech, as the output. However, for spoofed speech, it will generate very different output that amplifies the difference in the distribution compared to genuine speech. Therefore, they propose a genuinization transformer that uses genuine speech features with a convolutional neural network (CNN). The genuinization transformer is then used together with a Light Convolutional Neural Network (LCNN) system for the detection of synthetic speech attacks. It was trained on the ASVspoof2019 dataset with Constant Q Cepstral Coefficients (CQCC) and Linear Frequency Cepstral Coefficients (LFCC) as features, and achieved an EER of 4.07\% and a min-TDCF of 0.102. However, replay attacks were not done, and this model was not tested across corpora. This paper, \cite{Yangfeatures}, introduces an inversion module that derives four features: inverted constant-Q coefficients (ICQC), inverted constant-Q cepstral coefficients (ICQCC), constant-Q block coefficients (ICBC), inverted constant-Q linear block coefficient (ICLBC). The first two use conventional Discrete Cosine Transform (DCT) and the latter two use overlapped block transform. These features for synthetic speech detection are evaluated using the ASVspoof 2015, noisy ASVspoof 2015, and ASVspoof 2019 logical access databases, and achieve an EER of 7.77\% and a min-TDCF of 0.187. Suthokumar et al, in \cite{8682411}, compare how genuine and spoofed speech varies across different phonemes, and identifies certain phonemes (fricatives, nasals, stops, and pauses) which are more informative in the detection of replay attacks. The paper created four different fusion scoring methods to incorporate phonetic information using phoneme-specific models, and achieved an EER of 6.18\% on the ASVspoof2017 V2 dataset. However, this model has to be fused with a phenome independent model for better results. For logical scenarios, this paper uses feature extraction, a densely connected network, a squeeze and excitation block, and a feature fusion strategy. For physical scenarios, their method consists of feature extraction, a multi-scale residual network, an SE block and weighted average strategy. Both methods for logical and physical scenarios combine physical and physiological characteristics to resist spoofing attacks. \\Adversarial attacks may also pose a threat to ASV systems. According to \cite{AADefense}, ASV systems are prone to adversarial attacks, which may reduce the accuracy of such systems by up to 94\%. Attacks can range from simple Gaussian noise to more advanced attacks created in a targeted white-box setting. These perturbations, imperceptible to humans, may cause audio classification and ASV systems to fail completely.

\begin{table*}[!b]
\caption{\label{AACountermeasuresDefense}A Comparison of Adversarial Defensive Methods for Anti-Spoofing Models}
\centering
\begin{tabular}{p{1cm} | p{2.5cm} | p{2.5cm} | p{2cm} | p{2.8cm} | p{2.5cm} }
  \hline
 Paper & Technique & Dataset & Adversarial Testing & Key Performance & Limitations \\
  \hline
  \cite{wu2020defense} & Mockingjay with SSL & ASVspoof 2019 LA and LibriSpeech & FGSM and PGD & ~90\% accurate & limited to 2 attacks testing\\
  \hline
  \cite{wu2020defense2} & Spatial Smoothing and Adversarial Training & ASVspoof 2019 LA & PGD & 92.40\% (SENet), 98.60\% (VGG) &  Limited to 1 attack only\\
  \hline
\end{tabular}
\end{table*}

\begin{table*}[!b]
\caption{\label{AAASVDefense}Comparison of Adversarial Attack Defense for ASV Systems}
\centering
\begin{tabular}{ p{1cm} |p{2.8cm}|p{2.5cm}| p{2.8cm} |p{2.8cm}| p{2.5cm}}
  \hline
 Author & Technique & Dataset & Tested Adversarial Attacks & Best Evaluation Performance & Limitations \\
 \hline
 \cite{AADefense} & Adversarial Lipschitz Regularization & Librispeech & PGD, FGSM, Carlini, Wagner & 73\% adversarial accuracy & Fails with stronger attacks \\
 \hline
 \cite{wu2021spotting} & Neural Vocoders & Voxceleb1 and Voxceleb2 & BIM & AUC: 99.94 & Single Attack Testing \\
 \hline
 \cite{wu2021adversarial} & TERA models & Voxceleb1 & BIM & 22.94\% EER & Only tested on 1 type of attack \\
 \hline
 \cite{wu2021improving} & SSLMs & Voxceleb1 & BIM, FGSM, JSMA & R-Vector 14.59\% GenEER 
X-Vexctor 11.29\% GeEER
 & Degrades performance \\
 \hline
\end{tabular}
\end{table*}
Adversarial training is another way to reduce the effectiveness of an adversarial attack, as done by \cite{AADefense}. In this paper, adversarial training-- when adversarial samples are included in the training dataset-- and Adversarial Lipschitz Regularization (ALR) are used to reduce the impact of adversarial attacks. ALR is based on a function that ignores small changes in the input. Adversarial training is beneficial for defense when the attacks are similar to each other, but not all attacks will be similar. This method starts to fail when introduced to different types of attacks, some stronger than others, and this failure is evident when they are evaluated. 
The work of \cite{wu2021spotting} uses neural Vocoders to re-synthesize audio and finds differences between the ASV scores of an original audio sample and a re-synthesized audio sample. This method detects adversarial samples by first purifying the sample, i.e., removing adversarial noise while generating the genuine waveform with reduced distortion. This method is beneficial because it does not need to know the attack algorithm used. This model is tested purely on a Basic Iterative Method (BIM) adversarial attack, which extends an FGSM attack where FGSM is performed multiple times with a small step size. Adversarial attack types exhibit different effects, so judging performance on many types of attacks is difficult. Further, only one type of neural vocoder was used. Vocoders, when used to reconstruct phase information, inevitably introduce noise and distortion. The authors claim that this noise is actually beneficial, but this may not be entirely accurate. In addition, self-supervised learning-based models are more common when it comes to detecting and purifying adversarial samples. Self-supervised learning may be used  to reconstruct audio, which is beneficial when it comes to adversarial attacks on audio. 

The work of \cite{wu2021adversarial} uses self-supervised learning-based models for adversarial defense on ASV systems by using Transformer Encoder Representations from Alteration (TERA) models. The authors use self-supervised learning in a similar way to Mockingjay, which was a considerable influence on this paper. This method is beneficial because it does not require knowledge about how the adversarial samples were generated, as it purifies the samples. The authors use Gaussian, mean, and median filters for audio samples. Similarly, the work of \cite{wu2021improving} uses Self Supervised Learning Models (SSLMs) to remove superficial noise from the inputs, and reconstructs clean samples from the interrupted ones. They have two modules; one module purifies audio, while the other compares the original and purified samples in order to detect adversarial samples.

\section{Taxonomy of Voice Spoofing Countermeasures}
\begin{figure*}[t]
    \centering
    \includegraphics[width=13 cm]{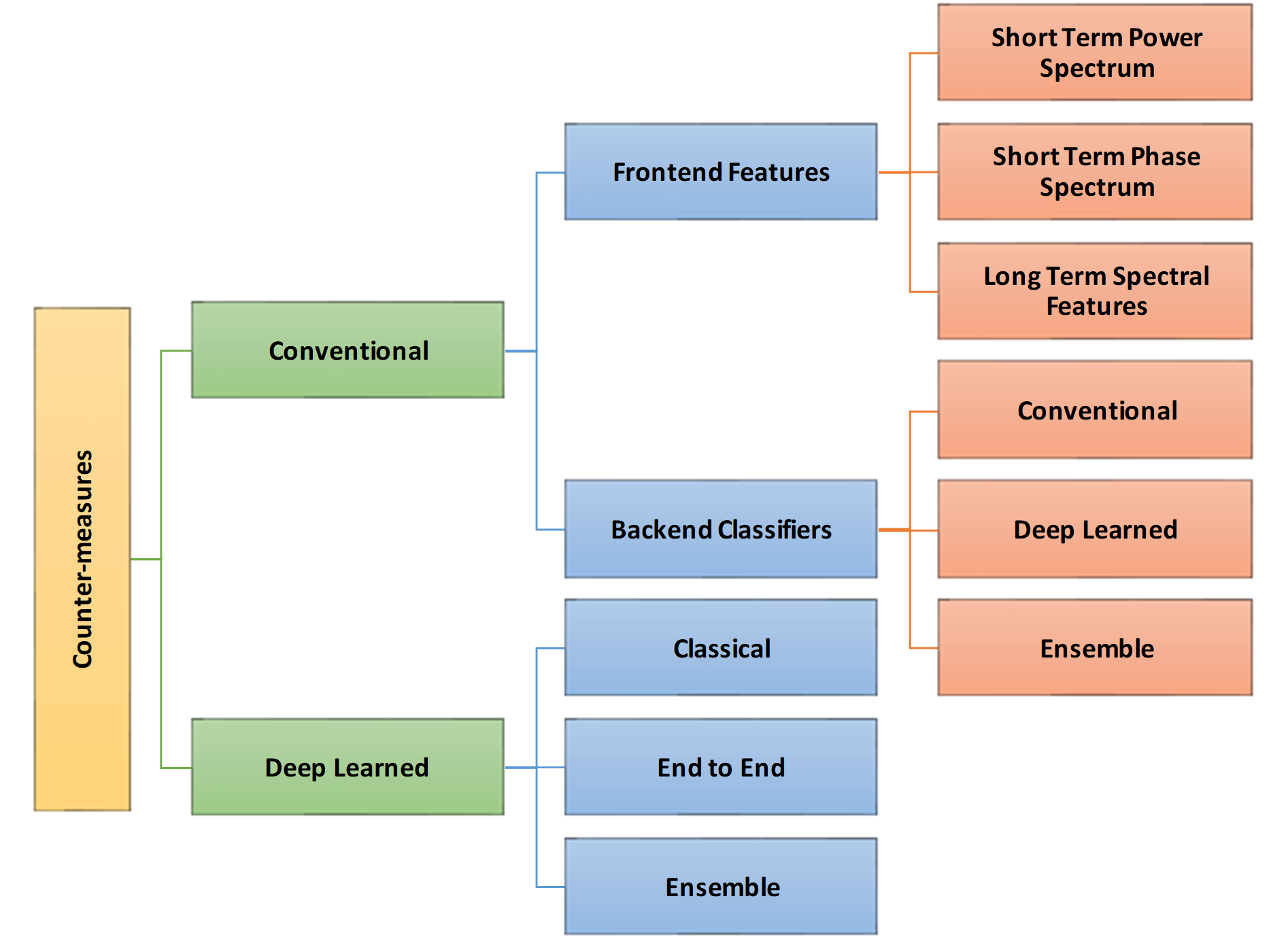}
    \caption{A detailed taxonomy of existing voice spoofing countermeasures. }
     \label{fig:Figure4}
\end{figure*}

This section provides a detailed analysis of existing state-of-the-art voice spoofing techniques adopted to detect spoofing attacks as well as countermeasures developed to combat those attacks. The countermeasure taxonomy illustrated in Figure~\ref{fig:Figure4} includes a comprehensive classification of the countermeasures. 

\subsection{Conventional Handcrafted Spoofing Countermeasures}
The countermeasures for audio spoof detection are classified into two categories: conventional handcrafted features and enhanced deep learning solutions. For voice spoofing detection, handcrafted features are commonly employed, i.e., MFCC \cite{zheng2001comparison}, CQCC \cite{Todisco2017ConstantQC}, and others, with conventional machine learning classifiers, e.g., Gaussian Mixture Models (GMMs) \cite{bond2001gmm}, Support Vector Machine (SVM) \cite{vishwanathan2002ssvm}, etc. 

In \cite{paul2015novel}, the author's novel acoustic features were derived from the frequency-warping and block transformation of filter bank log energies. A two-level Mel and speech-based wrapping was used with an Overlapped Block Transformation (OBT) to extract the conventional and inverted classification features. Although the proposed features achieved a 0.99 classification accuracy on the ASVspoof 2015 corpus development section, they were only applicable when the spoofing attacks were known in advance. Aside from this, the performance of the proposed approach was not evaluated and reported in the presence of unknown adversarial attacks. Further, the approach was not validated across corpora due to a lack of open-source datasets. In the ASVspoof2015 challenge, the unified articles concentrate on countermeasures for speech synthesis and voice conversion spoofing attacks. Conversely, replay attacks pose the greatest threat to ASV systems. These attacks involve the playback of recordings, acquired from registered speakers, in order to generate false authentication tokens, and may be efficiently carried out using common consumer devices.

 ASVspoof2017, the second in the series of challenges, focused on the development of replay attack countermeasures. In \cite{kinnuneninproceedings}, Kinnunen et al. mainly followed this line of thinking, utilizing front-end features, i.e., CQCC, MFCC, PLP, RFCC, Inverted Mel Frequency Cepstrum Coefficients (IMFCC), LPCC, SCMC, SSFC, and HPCC, with a baseline GMM-UBM classifier. The performance of these countermeasures was evaluated on the ASVspoof2017 dataset, where the proposed systems achieved an EER of 31.5\% for genuine vs. replay and 1.8\% EER for genuine vs. zero-effort imposters. In addition, the average detection EER of all main submissions was 25.91\%, whereas the best single system result had an average detection EER of 6.73\%. When these results were compared to those from the prior challenge (ASVspoof2015), the detection of replay attacks was found to be more complex than detecting speech synthesis and voice conversion spoofing attacks. 
In \cite{sahidullah2016integrated}, Sahidullah et al. present the first comparative evaluation of six countermeasures (CMs) and their integration with automatic speaker verification using the ASVspoof 2015 dataset. The CMs contain MFCCs, IMFCCs, LFCCs, Cochlear Filter Cepstral Coefficients (CFCCs), CQCCs, and Gammatone Frequency Cepstral Coefficients (GFCCs) as front-end cepstral features, which are then coupled with GMM-UBM and i-Vector classifiers for back-end speaker classification. According to the results presented in this paper, the CM using CQCC features and the fusion of all six CMs has the greatest potential to detect spoofing attacks in the context of spoofing attacks. More significantly, the fusion of CM and CQCCs achieves the lowest EER of 0.02\%.

 Cascading integration of ASV and CMs significantly decreases the False Acceptance Rate (FAR). Although the significance of CQCC features is demonstrated in this study, the presented system is not evaluated against other forms of spoofing attacks, e.g., replay, deepfake, adversarial, or others. Further, countermeasures are not tested cross-corpus. The results of \cite{sahidullah2016integrated} demonstrate that an ensemble of acoustic features, together with machine learning models, can produce more accurate results than individual classifiers. Consequently, in \cite{ji2017ensemble}, the author presents an ensemble of the acoustic features of CQCCs, along with other classical features, i.e., MFCCs and Perceptual Linear Predictive (PLP) features. The author proposes an ensemble classifier set that contains numerous GMMs, i.e., Gradient Super Vector-Boosting Decision Tree (GSV-GBDT) and GSV-Random Forest (GSV-RF) classifiers. The experimental results show that the presented ensemble system significantly outperforms the baseline GSV-SVM system. Specifically, using a baseline GSV-SVM classifier, this approach achieves an EER of 10.4\% on CQCC features, 27.4\% on MFCC features, and 37.0\% on PLP features. In contrast, a minimal EER of 9.5\% is achieved by employing the ensemble model. Similarly to \cite{paul2015novel}, the biggest limitation of \cite{ji2017ensemble} is that it has only been tested against a single replay attack. Other types of spoofing attacks, i.e., speech synthesis, voice conversion, and the most current deepfake and adversarial attacks, are not reviewed, and there is no cross-corpus examination of the techniques. In \cite{balamurali2019}, the authors examine the robust audio features, comprised of handcrafted and auto-encoder-based learned features, to identify replay spoofing attacks. The handcrafted features employed in this study are CQCCs, LPCCs, IMFCCs, Rectangular Filter Cepstral Coefficients (RFCCs), Sub-band Centroid Frequency Coefficients (SCFCs), and Sub-band Centroid Magnitude Coefficients (SCMC), as well as spectrogram features. The paper uses an auto-encoder to learn a dense representation of all of the features. A conventional GMM, along with a Universal Background Model (UBM), is used as the baseline system to examine the performance of the handcrafted and encoder-based features. The integrated fused models, based on existing audio and machine-learned features, achieve comparable results, with an identical error rate (EER) of 12.0\%. In particular, the handcrafted CQCC features outperform all other features, with an EER of 17.5\%. The best encoder-based feature observed was a spectrogram with a minimum EER of 20.2\%. Although the coupling of handcrafted features with an auto-encoder-based system surpasses the state-of-the-art, the presented system is only tested against replay attacks. In the presence of other types of spoofing attacks, e.g., voice conversion and adversarial attacks, the system's performance tends to vary.
 
Rather than using standard standalone short-term power spectrum coefficients, i.e., MFCCs, LFCCs, CQCCs, etc., to identify the spoofing attacks, the authors of \cite{tapkir2018significance} introduced phase-based Teager Energy Operator (TEO) features. It was discovered that the TEO phase features gave information that is complementary to the information provided by the more commonly used CQCC, MFCC, and LFCC feature sets. Although the TEO phase features were unable to perform well alone, fusion with traditional features enhanced the accuracy of the spoofing detection system. The results demonstrated that the standalone spoof detection systems, developed with the TEO phase, MFCC, and LFCC, achieved an EER of 31.34\%, 34.02\%, and 16.80\%, respectively, whereas, when the TEO phase feature set was fused with the CQCC, MFCC, and LFCC feature sets, the EER was lowered by 0.18\%, 2.74\%, and 1.41\%, respectively. This improvement in system performance showed the influence of TEO phase information on the spoof detection system. However, the provided solution was only stated to be resistant to replay spoofing attacks, and had no cross-corpus validation.

In accordance with \cite{tapkir2018significance}, the authors of \cite{novoselov2016stc} present phase spectrum and multi-resolution wavelet features, in addition to the commonly used front-end MFCC features. This study combines MFCCs with Mel-Frequency Principal Coefficients (MFPCs), CosPhase Principal Coefficients (CosPhasePCs), and Mel Wavelet Packet Coefficients (MWPCs) to provide a reliable and robust defense against spoofing attacks. The experimental results indicate that applying principal component analysis (PCA) to MFCCs results in a considerable EER improvement over MFPC features for all spoofing methods. However, the MFCC features prove inferior in comparison with other front end features. In contrast, the implementation of CosPhasePC modestly decreases EER in comparison to the MFPC features. Although the multi-resolution wavelet transform features outperform the state-of-the-art, achieving 0.05\% EER for all known attacks, the proposed framework has not been tested and reported cross-corpus and against unknown attacks. Accordingly, following the significant performance of phase-oriented features, the author sheds light on the importance of acoustic front-end features and introduces a novel detection mechanism by modeling replayed speech as a convolution of original speech \cite{tapkir2018novel}. Also in \cite{tapkir2018novel}, the author proposes a novel feature set, Magnitude-based Spectral Root Cepstral Coefficients (MSRCC) and Phase-based Spectral Root Cepstral Coefficients (PSRCC), which outperforms the baseline system (CQCC) and provides a 29.18\% EER on the evaluation set of the ASVspoof 2017 challenge database. With the GMM back-end classifier, the front-end features MSRCC and PSRCC respectively produce 18.61\% and 24.35\% EER. Conversely, with convolutional neural network (CNN) back-end classifiers, MSRCC and PSRCC obtain 24.50\% and 26.81\% EER, respectively. In addition, the score-level fusion of MSRCC and PSRCC results in 10.65\% and 17.76\% EER using GMM and CNN classifiers, respectively. These findings suggest that the proposed feature sets of MSRCC and PSRCC capture complementary information. However, the cross-corpus examination was not reported. 

The research in  \cite{cai2017countermeasures} shows the effectiveness of augmenting genuine training data in the simulation of replay spoofing attacks. The author presents replay spoofing countermeasure systems that improve the CQCC-GMM baseline with score level fusion. Instead of using CQCC, the author used spectrograms as inputs to analyze end-to-end feature representations. Finally, the author replaced the subsequent GMM classifier with a Fully-connected Deep Neural Network (FDNN) and a Bidirectional Long-Short-Term Memory neural network (Bi-LSTM). The results of the experiments show that this data augmentation technique can significantly increase the system's performance. In particular, the baseline CQCC-GMM model achieved an EER of 22.29\%, while the DA-CQCC-GMM model obtained an EER of 19.18\% and the fused system achieved an EER of 16.39\%. Although the end-to-end FDNN and Bi-LSTM-based systems perform well for replay detection, other sorts of current deepfake and adversarial attacks may have led the system to perform inadequately. Furthermore, the provided system was not evaluated in the presence of unknown attacks and audio recordings across corpora. 

Yang et al. \cite{YANG201930} developed a Low-Frequency Frame-wise Normalization (LFFN) technique to capture replay distortions. LFFN was combined with Constant-Q Transforms (CQT) to extract two features: Constant-Q Normalization Segmentation Coefficients (CQNSC) and Constant-Q Normalization Cepstral Coefficients (CQNCC). This approach performed well on the ASVspoof 2017 version 2.0 dataset, with an EER of 10.63\% on CQNSC and 10.31\% on CQNCC features. The novel Acoustic Ternary Patterns-Gammatone Cepstral Coefficient (ATP-GTCC) feature was proposed in \cite{Malik2020ALight} to help develop a lightweight model for single and multi-order replay attack detection. An SVM classifier was used to better capture the harmonic distortions found in multi-order replay samples, while ATP-GTCC was employed as a front end feature. This model achieved an EER of 0.6\% and 1\% on the Voice Spoofing Detection Corpus (VSDC) and the ASVspoof 2019 dataset, respectively. This system exceeded earlier state-of-the-art techniques in terms of replay detection performance and efficiency. However, the given approach is limited to single-order and multi-order replay attacks, and has not been tested against other types of attacks, including voice conversion, deepfake, and adversarial attacks. 

In \cite{7041636}, Wu et al. use a common benchmarking database to analyze the vulnerability of text-dependent speaker verification systems to the replay attack and propose an anti-spoofing technique to protect the ASV systems. The main idea behind this spoofing detection approach is to use a similarity score to determine whether the provided sample is identical to any previously stored voice samples. A back-end HMM classifier is trained using the front-end MFCC, LPCC, and spectro-temporal features. Experiments on the RSR2015 database reveal that in the presence of a replay attack, the EER and FAR, which were identical at 2.92\%, increased to 25.56\% and 78.36\%, respectively, proving the speaker verification system's vulnerability to replay attacks. In the presence of replay spoofing attacks, the proposed spoofing countermeasure lowers the FARs from 78.36\% and 73.14\% to 0.06\% and 0.01\% for male and female samples, respectively. In \cite{9445238}, the authors use a feature fusion of GTCC, MFCC, Spectral Flux, and Spectral Centroid for input audio presentation. This countermeasure successfully detects multiple types of logical access attacks and classifies the cloning algorithms used to produce the synthetic speeches. It achieves an EER of 3.05\%, compared to baseline models that achieve an EER ranging between 5.06\% and 9.57\%. The authors of \cite{9638512} present a voice spoofing countermeasure using ELTP-LFCC features and a Deep Bidirectional LSTM to combat TTS synthesis and converted voice samples in LA attacks. In this paper, ELTP is further fused with LFCC to better capture the characteristics of the vocal tract speech dynamics of both bonafide voice and cloning algorithm artifacts. On the diverse ASVspoof 2019-LA dataset, performance evaluation yields an EER of 0.74\% and a min-TDCF of 0.008\%. However, the presented system is only tested on one dataset.

Rigorous experimentation was performed to illustrate the significance of the proposed countermeasure for detecting LA-based voice spoofing attacks. In the literature, frequency or amplitude modulation-based features were also explored for voice spoofing detection. Gunendradasan et al. \cite{gunedradasaninproceedings} used Spectral Centroid Deviation (SCD) features to develop a replay attack detection system. The Spectral Centroid Frequency (SCF) and Spectral Centroid Magnitude Coefficient (SCMC) features were extracted from the same front-end as SCD and used as complementary features. These features were used to train a GMM classifier for replay attack detection. This method was evaluated on the ASVspoof2017 dataset and provided a 60\% improvement in EER as compared to the ASVspoof CQCC baseline model.

Existing methods have also used various handcrafted features to address both physical (replay) and logical access (clone, voice conversion) spoofing. Monteiro et al. \cite{monteiro2020ensemble} proposed a system that can detect both logical and physical access audio attacks against ASVs via an ensemble of three different models. The baseline systems of LFCC-GMM and CQCC-GMM obtained an EER of 8.09\% and 9.57\%, respectively, on the evaluation set. Although this system achieved an EER of 9.87\% for voice cloning attacks, for replay attacks, the baseline systems achieved an EER of 11.04\% and 13.54\%, whereas the proposed system achieved an EER of 1.74\%. Zhizheng et al. \cite{7858696} proposed an ASV spoofing and countermeasures initiative with more focus on speech synthesis and voice conversion spoofing attacks (presentation attacks). The authors also described the post-evaluation results of the ASV systems achieved by the fusion of several spectral features, i.e., MFCC, CFCCIF, and MFPC, as well as the back-end classifiers, such as GMM, SVM, SVM-Fusion, and SVM with i-Vectors. 

Some methods also exploit high-frequency components in order to capture the traits of bonafide and spoofed signals. The work in \cite{witkowski2017audio} captured the high-frequency content by using the inverted-MFCC (IMFCC), LPCC, LPCCres, CQCC, MFCC, and Cepstrum features. This hybrid feature representation was then used to train a GMM for the classification of bonafide and spoofed samples. Due to the increased feature computation cost, this method was not suitable for local deployment on resource-constrained voice-controlled systems. This paper, \cite{8682771}, also highlighted the idea of analyzing the high-frequency band for replay spoof detection. For this purpose, transmission line cochlea amplitude modulation and transmission line cochlea frequency modulation features were employed to train a GMM for replay attack detection. This method was evaluated on the ASVspoof2017 dataset and achieved an EER of 7.59\%. Although this method \cite{8682771} performed only slightly better than the baseline model, amplitude modulation-based features have a high computational cost because they take more than twice the amplitude frequency to modulate the signal.

In the literature, non-voice segments have also been explored to capture the distortions of playback speech. Saranya et al., in \cite{8724469}, analyzed the channel and reverberation information from non-voiced segments of the input audio for replay spoofing detection. A voice activity detector was used to determine the non-voiced segments, then a hybrid feature vector comprised of CQCC, MFCC, and Mel-Filter bank-Slope was employed to capture the remnant vocal tract information in the non-voiced segments of the input audio. These features were later used to train a GMM for the classification of bonafide and replayed audio samples. The performance of this method was evaluated on the ASVspoof2017 dataset, and showed an improvement of 37\% in EER over the baseline method. In the work of \cite{Yangfeatures}, the authors introduced an inversion module to derive four features: ICQC, ICQCC, ICBC, and ICLBC. Two of them used conventional DCT, and the other two used overlapped block transforms. These features for synthetic speech detection were evaluated using the ASVspoof 2015, noisy ASVspoof 2015, and ASVspoof 2019 logical access datasets, and achieved an EER of 7.77\% and a min-tDCF of 0.187. Despite the fact that they were tested on multiple datasets, due to aadvances in spoofing attacks, i.e., adversarial or unknown attacks, they may not perform well in a real-world scenario.
\subsection{Discussion}
We performed a survey to determine the most frequently used features for spoofing audio and speech countermeasures. Each handcrafted feature was carefully designed and has its own dependencies to differentiate between genuine and spoofed voice samples. Despite the fact that each spoofing attack adds a distinct distortion to the speech sample in order to mislead an ASV system, some recently published crafted features worked very well in the classification of physical and logical attacks. Although the reported hand-crafted features have been evaluated on several datasets, including ASVspoof2015, ASVspoof2017, ASVspoof2019, and others, they are dependent on the data source used to train and classify the spoofed and bonafide speech samples. There is a significant need for a standardized, unified dataset to train and validate the accuracy of the published feature sets in order to choose the best among them. Furthermore, the existing features have only been tested on training datasets so far, with no cross-corpus assessment. Consequently, the list of features we have chosen covers the bulk of the power, frequency, duration, amplitude, and phase spectrum in order to distinguish between bonafide and spoofed voice samples. Moreover, we offer a cross-corpus examination of the features and test performance on standard datasets. The experiment and results section provides the details of the performance analysis on the selected speech samples. 

\subsection{Deep learning based Countermeasures} 
Recent years have witnessed a rise in the use of deep learning approaches to prevent audio spoofing attacks. However, despite the benefits of handcrafted features, these approaches were computationally complex and required powerful computing resources. Inspired by the great success of deep learning in automatic speech recognition, deep neural network (DNN) based systems were developed for spoofing detection for the first time in 2015. In \cite{chen2015robust}, a novel, simple model for detecting spoofing attacks on a speaker verification system was developed. The presented model was used to extract key features from audio samples and construct compact, abstract, and resilient deep data representations. A spoofing-discriminant network was used in the training of spoofing algorithms. The proposed network then computed the s-vector, which is the utterance level average of the final hidden layers. Finally, Mahalanobis distance, along with normalization, was used, in conjunction with computed s-vectors, to detect spoofing attacks. This model achieved the 3rd position in the first spoofing detection challenge, ASVspoof 2015. In particular, the proposed model attained an overall minimum EER of 2.281\%. Specifically, the presented system obtained an EER of 0.046\% for known and 4.516\% for unknown replay attacks. Although the proposed system performed well in the ASVspoof 2015 competition, it has not been examined across corpora. Furthermore, when spoofing techniques other than the replay attack are introduced into the system, the performance of the proposed technique is taken into account. 

\par Existing methods also employed handcrafted features to train deep neural networks for spoofing detection. In \cite{Nagarsheth2017ReplayAD}, high-frequency regions were analyzed by proposing the High Frequency Cepstral Coefficients (HFCC) feature and using a back-end DNN model to classify bonafide and spoofed (replay) audio. When comparing HFCC features to CQCC features using a baseline GMM back-end, HFCC outperformed CQCC in the development and evaluation sets of ASVspoof2017. HFCC received an EER of 5.9\% in the development set and 23.9\% in the evaluation set, while CQCC received 11.0\% EER in the development set and 24.7\% in the evaluation set. An attentive filtering network system was proposed in \cite{lai2019attentive} to detect replay attacks against ASV systems, and the proposed method enhanced feature representations in both the frequency and time domains. This system obtained an EER of 6.09\% and 8.54\% on the development and evaluation sets of ASVspoof 2017, respectively, on a system comprised of two Attentive Filtering models (one using sigmoid and the other using softmax as their activation functions). This method achieved better results than the CQCC-GMM baseline model, which obtained an EER of 12.08\% and 29.35\%. However, the combination of feature extraction and the model's architecture required significant computational resources.

In \cite{huang2020audio}, segment-based linear filter bank features, along with an attention-enhanced DenseNet-BiLSTM model, were proposed for replay attack detection. These features were extracted from the silent segments of the audio signal. Triangle filters were used to examine noise in the high frequency bands \(3-8 kHz\). The baseline system of CQCC-GMM obtained an EER of 2.36\% on the development set and an EER of 8.42\% on the evaluation set of ASVspoof2017, while features of this method, when used with a GMM, obtained an EER of 1.8\% and 7.92\% on the development and evaluation sets, respectively. One limitation of this method was the sensitivity of the signal-to-noise ratio during segmentation of the silence and voiced components, where low SNR resulted in false segmentation.

Even though the work of \cite{wuCNN} uses LPS features, the enhanced modification was before passing them on to an LCNN. The authors create a model that fits the distribution of genuine speech, i.e., one that takes genuine speech as the input and generates genuine speech as the output. However, if the speech is spoofed, the output will be very different. They propose a genuinization transformer that uses genuine speech features with a convolutional neural network (CNN). The genuinization transformer is then used with an LCNN system for the detection of synthetic speech attacks. The model achieves an EER of 4.07\% and a min-tDCF of 0.102 on the ASVspoof2019 dataset, using CQCC and LFCC features. However, this method has not been tested across corpora. The work of \cite{Ma} proposes a Conv1D Resblock with a residual connection, which allows the model to learn a better feature representation from raw waveforms. They find that feeding a raw waveform directly into a neural network is adequate. Only the ASVspoof2019 dataset is used, and it achieves an EER of 2.98\% and a min-tDCF of 0.082. The work of \cite{Tak} is an extension of a previously proposed GAT-ST and uses a raw waveform. This captures discriminative cues in both the spectral and temporal domains. The proposed RawGAT-ST model uses a one-dimensional convolution layer to ingest raw audio. This end-to-end architecture uses feature representation learning and a GAT that learns the relationships between cues at different sub-band and temporal intervals. This model uses the raw waveforms from the ASVspoof 2019 dataset and achieves an EER of 1.06\% and a min-tDCF of 0.0335. However, this model is only tested on one dataset. Yang and Das, in \cite{YANG2020107017}, propose a new feature that can capture discriminative information between natural and spoofed speech. The proposed feature, eCQCC-STSSI, uses CQT to transform speech from the time domain to the frequency domain. A DCT is used to decorrelate the feature dimensions and concentrate the energy of the logarithmic octave power spectrum and logarithmic linear power spectrum, respectively. The two DCT outputs are concatenated to form the eCQCC feature vectors. Even though this feature was tested on both ASVspoof2015 and ASVspoof2017, it has not been tested on other datasets and may fail to properly extract features.
The authors of \cite{9152815} concatenate four sub-features on the ASVspoof2019 dataset. These features, Short-Term Spectral Statistics Information (STSSI), Octave-band Principal Information (OPI), Full-band Principal Information (FPI), and Magnitude-Phase Energy Information (MPEI), are fused to generate the delta acceleration coefficients as features for spoofing detection such as CQSPIC, CQEPIC, and CESPIC. The fusion of the features achieves an EER of 7.63\% and a min-TDCF of 0.178, but has been tested in many scenarios. In \cite{8865624}, the authors propose a heuristic feature extraction method based on Multi Level Transform (MLT), which extracts valuable information from the octave power spectrum for spoofing attack detection. It relies on MLT to extract relevant information from previous DCT results. The authors apply it to the ASVspoof2015 and ASVspoof2017v2 datasets, and achieve an EER of 14.45. However, it is only tested on two domains. This paper, \cite{8682411}, compares genuine and spoofed speech across different phonemes and shows that specific phonemes (fricatives, nasals, stops, and pauses) are more informative in the detection of replay attacks. The paper creates four different fusion scoring methods to incorporate phonetic information using phoneme-specific models. This method is tested on the ASVspoof2017 V2 dataset, and achieves an EER of 6.18\%. However, it has to be fused with a phenome-independent model for the best results. The work of \cite{8737422}, which also looks at sound characteristics, creates Voice-Pop, which identifies a live user by detecting the pop noise naturally incurred by a user breathing while speaking close to the microphone. The authors use their own dataset with GFCC features and achieve an EER of 5.4\%. However, this method was only tested in one domain. 
A few solutions use lightweight deep learning systems to detect replay spoofing. A lightweight CNN \cite{lavrentyeva2019stc}, based on Maximal Feature-Map (MFM) activation, is used to detect replay attacks in \cite{wu2018light}. MFM is able to minimize the dimensionality by using the most relevant features for classification. This method, \cite{wu2018light}, is extended in \cite{balamurali2019toward} to investigate the efficacy of angular margin-based softmax activation in training a light CNN for cloning and replay spoof detection.

\subsection{End to End Countermeasures} 
End-to-end techniques have been shown to be effective in a variety of domains. However, relatively little work has been done in the domain of voice spoofing countermeasures to date. In \cite{endtoend2017}, an end-to-end system was proposed to detect replay attacks against ASVs. This system used a combination of a CNN, LSTM, and DNN to take in raw audio as input and classify the audio into genuine, synthetic/cloned, or replay. This model was evaluated on the ASVspoof 2015 and BTAS 2016 datasets and achieved a half total error rate (HTER) of 1.56\%, compared to the GMM-PLP-39 baseline model of 2.96\%. This system may be enhanced in the feature extraction stage by employing better time and frequency filtering networks. Recently, in \cite{Jung}, the author proposes a novel heterogeneous stacking graph attention layer that models artifacts spanning heterogeneous temporal and spectral domains with a heterogeneous attention mechanism and a stack node. In concert with a new max graph operation that involves a competitive mechanism and an extended readout scheme, their approach, named ASSIST, achieves an EER of 0.83\% and a min-tDCF of 0.0275 on ASVspoof2021. However, the proposed method is only tested on one dataset. 

In \cite{Tak2}, a graph attention network (GATs) is proposed that works by applying a self-attention mechanism to GNNs and modeling graph-structured data. Each node in the graph is weighted according to its relevance to other nodes. GATs can be used to model a specific sub-band or temporal segment using high-level representations extracted from deep residual networks. It achieves a min-tDCF of 0.0089 on the ASVspoof2019 dataset. However, GAT operates on filter bank outputs, not on the waveform. It is only tested on one dataset. 

\subsection{Discussion}
The majority of these papers do not test across multiple datasets to establish the generalizability of the solution, perhaps due to the lack of publicly available datasets at the time of publication. However, there are numerous datasets available for testing right now, and a detailed analysis of cross-corpus evaluation of the solutions is presented in Table V. This analysis shows that there are only two articles \cite{Todisco2017ConstantQC,baumann2021voice}that have performed the cross-corpus evaluation to date. However, there are some research gaps in the work of these two papers, and they do not perform a thorough analysis of cutting-edge techniques. Some of these papers use raw wave forms, which can save time in feature extraction but need to be verified under controlled conditions. Also, the depth of each model varies, which can increase accuracy on test sets but take longer to train. In terms of deployment, ideally models need to be lightweight enough to be run on smart speakers. New machine learning techniques for the audio domain can be adapted from the video domain, which is generally more developed. In a world that changes frequently, the models need to be able to be adapted to new challenges, and some of these models are designed for a specific task. New models need to be versatile enough for various conditions. Additionally, these models need to be unbiased, an aspect that has not been tested.

\begin{table}[t]
\label{tab:authorcorpus}
\caption{Cross corpus evaluations of existing methods}
\setlength{\tabcolsep}{0.7\tabcolsep}
\begin{tabular}{p{0.8cm}|p{4cm}|p{2.2cm}}
\hline
Author & Corpus & Cross-Corpus\\
  \hline  \hline
  \cite{9638512} & ASVspoof 2019 LA & No \\
  \cite{9445238} & ASVspoof 2019 LA & No \\
  \cite{10.1007/978-3-319-67934-1_9} & chsc2011 \& GTZAN & No \\
  \cite{kua2010investigationOS} & NIST 2001 \& 2006 \& SRE  & No \\
  \cite{saratxagainproceedings} & In-house dataset & No \\
  \cite{gunedradasaninproceedings} & ASVspoof 2017 v1.0 & No \\
  \cite{kinnuneninproceedings} & ASVspoof 2017 & No \\
  \cite{Nagarsheth2017ReplayAD} & ASVspoof 2017 & No \\
  \cite{7858696} & ASVspoof 2015 & No \\
  \cite{murthy2003modified} & SPINE2000 & No \\
  \cite{rajan2013using} &  NIST 2010 evaluation & No \\
  \cite{8724469} & ASVspoof 2017 dataset & No \\
  \cite{8682771} & ASV spoof 2017 & No \\
  \cite{7041636} & Part I-RSR2015 corpus  & No \\
  \cite{YANG201930} & ASVspoof 2017 v2.0 & No \\
  \cite{cai2017countermeasures} & ASVspoof 2017 & No \\
  \cite{9107388} & VSDC \& ASVspoof 2019 & No \\
  \cite{chettri2020deep} & ASVspoof 2017, 2019 PA & No \\
  \cite{chen17_interspeech} & ASVspoof 2017 & No \\
  \cite{Todisco2017ConstantQC} & ASVspoof 2015, 2017 \& RedDots & Yes \\
 \cite{yang2020} & ASVspoof 2017 V2.0 & No \\
 \cite{wu2020defense2} & ASVspoof 2019 LA & No \\
 \cite{wu2020defense} & ASVspoof 2019 LA & No \\
 \cite{kinnunen2018t} & ASVspoof 2015,2017 & No \\
 \cite{witkowski2017audio} & ASVspoof 2017 & No \\
 \cite{wang2020asvspoof} & ASVspoof 2019 & No \\
 \cite{baumann2021voice} & ASVspoof 2017,2019 \& ReMASC & Yes \\
 \cite{endtoend2017} & BTAS2016 dataset & No \\
 \cite{monteiro2020ensemble} & ASVspoof 2019  & No \\
 \cite{tapkir2018significance} & ASVspoof 2017 & No \\
 \cite{ji2017ensemble} & ASVspoof 2017 & No \\
 \cite{sahidullah2015comparison} & ASVspoof 2015 & No \\
 \cite{naika2018overview} & VoxForge & No \\
 \cite{tapkir2018novel} & ASVspoof 2017 & No \\
 \cite{huang2020audio} & BTAS2016 \& ASVspoof2017 & No \\
 \cite{lai2019attentive} & ASVspoof 2017 & No \\
 \hline
\end{tabular}
\end{table}

\subsection{Unified voice spoofing countermeasures}
By and large, existing countermeasures for voice spoofing attacks exclusively target one type of attack (e.g., voice replay or speech synthesis), and only recently few works have been reported that utilize a unified approach, capable of dealing with multiple types of voice spoofing attack. In \cite{javed2022voice}, the authors introduce a novel ATCoP feature descriptor for the detection of voice presentation attacks, capable of recognizing both LA and PA attacks. Although a good all-around solution, this unified anti-spoofing technique is the most effective at detecting single-and multi-order replay attacks. The experimental results show that the presented approach is effective when tested on four distinct datasets, with either replay or clone forgeries. Although this method is evaluated against several datasets, the success of the given ATCoP descriptors is not reported for DF attacks. In another work \cite{javed2021towards}, the authors recognized a novel hybrid voice spoofing attack, a cloned replay, which may also be used to spoof an ASV system. A cloned replay attack duplicates the audio of a target speaker and transmits it to an ASV system. The author establishes the basis for a spoofing countermeasure capable of detecting multi-order replay, cloning, and cloned-replay attacks through the use of the proposed ATP-GTCC features. This approach is capable of identifying state-of-the-art (SOTA) voice spoofing attacks with a unified solution.

Rostami and Homayounpour et al. \cite{rostami2021efficient} proposed an effective attention branch network (EABN) for detecting LA and PA attacks. The provided technique \cite{rostami2021efficient} achieves EERs of 0.86\% and 1.89\%, and min-TDCFs of 0.02\% and 0.50\%,  against PA and LA attacks, respectively. Despite outperforming SOTA approaches on LA and PA attacks, \cite{rostami2021efficient} was not tested on deepfake attacks. In contrast, in \cite{lai2019assert}, SENet and ResNet were combined with statistical pooling to handle anti-spoofing with deeper and faster-trained DNNs. It consisted of a SENet, a median-standard ResNet, a dilated ResNet, and an attentive filtering network. A GNN back end classifier was implemented using CQCC and LFCC features. It obtained an EER of 0.59\% for a PA and 6.7\% against LA attacks. Because the data was obtained directly from the ASVspoof2019 dataset with no data augmentation or cross-corpus dataset, the proposed method's performance was not checked against DF attacks and may decline in a real-world scenario. In \cite{chen21_asvspoof} the authors presented Emphasized Channel Attention, Propagation, and Aggregation Time-Delayed Neural Networks (ECAPA-TDNN) as their primary model. The authors' intention was to tackle the issue of channel variability by employing an acoustic simulator in order to enhance the original datasets with transmission codecs, compression codecs, and convolutional impulse responses. The presented method attained an EER of 5.46\% and a min-tDCF of 0.3094 in the 2021 LA task and an EER of 20.33\% on the DF task. Although the presented work prevents several types of attacks, it needs testing and reporting against PA attacks. To the best of our knowledge, no countermeasures have been reported for all four types of voice spoofing attacks, i.e., LA, PA, DF, and cloned replay attacks. However, the community's attention has lately switched to spoofing-aware ASV systems, as mentioned in the section below. New integrated spoofing technologies have been presented that conducted speech spoofing and automated speaker verification simultaneously, in addition to unified techniques for cutting-edge spoofing attacks. 

\subsection{Integrated spoofing-aware ASV systems}

The research community for secure voice-enabled systems is currently focused on integrating research efforts on speaker verification and anti-spoofing. Unlike existing anti-spoofing systems, which focus on independently streamlined spoofing detection, speaker verification may also be embedded in order to build a integrated system. A solution is considered as an Integrated spoofing-aware ASV systems if it is an integrated and optimized system that protects against spoofing attacks while also performing speaker verification. The first paper in this regard is \cite{jung2022sasv}, in which the authors propose the spoofing-aware speaker verification (SASV) challenge, which integrates speaker verification and anti-spoofing. In this challenge, the organizers encourage the development of integrated SASV systems that use new metrics to evaluate joint model performance by releasing official protocols and baseline models. The authors extend speaker verification by including spoofed trials in addition to the standard set of target and impostor trials. Unlike the existing ASVspoof challenge, which focuses on separate spoofing detection and speaker verification systems, SASV aims to develop jointly optimized secure ASV solutions. Open-source, pre-trained spoofing detection and speaker verification models are used in two baseline SASV solutions. Participants have free access to both models and baselines, which can be used to develop back end fusion approaches or end-to-end solutions. The top performing system reduces the equal error rate of a conventional speaker verification system from 23.83\% to 0.13\% when tested with target, bonafide non-target, and spoofed non-target trials. The SASV challenge results demonstrate the dependability of today's cutting-edge approaches to spoofing detection and speaker verification. In this paper, \cite{zhang2022probabilistic}, Zhang et al., develop a probabilistic framework for combining the ASV and CM subsystem scores. In addition to the probabilistic framework, the authors propose direct inference and fine-tuning strategies, based on the framework, to predict the SASV score. Surprisingly, these strategies reduced the SASV EER of the baseline to 1.53\% in the official SASV challenge evaluation trials. The author validates the efficacy of the proposed modules through ablation studies and provides insights through score distribution analysis. 

The author of \cite{liu2022spoofing}, expresses concern about the improvement of spoofing robustness in automatic speaker verification (ASV) systems in the absence of a separate countermeasure module. To address this issue, the ASVspoof 2019 baseline model is used, in accordance with the back-end machine learning classifier's probabilistic linear discriminant analysis (PLDA). Three unsupervised domain adaptation techniques are used to optimize the back-end using audio data from the ASVspoof 2019 dataset training partition. The results show significant improvement in both the logical and physical access scenarios, particularly in the latter, where the system is attacked by replayed audio, with maximum relative improvements of 36.1\% and 5.3\% in bonafide and spoofed cases, respectively. However, it is observed that absolute error rates on spoof trials remain too high on spoofing attacks. This demonstrates the challenge of making a conventional speaker embedding extractor with a PLDA back-end work on a mix of bonafide and spoofed data.

Over the last few years, research has improved the performance of ASVs and countermeasure systems, resulting in low EERs for each system. However, research on the joint optimization of both systems is still in its early stages. This paper, \cite{teng2022sa}, proposes a Spoof-Aggregated-SASV (SA-SASV), an ensemble-free, end-to-end solution for developing an SASV system with multi-task classifiers. An SA-SASV system is further optimized by multiple losses and more flexible training set requirements. The proposed system is trained using the ASVspoof2019-LA dataset. SA-SASV EER results show that training in complete automatic speaker verification and countermeasure datasets can improve model performance even further. The results show that the SA-SASV feature space outperforms previously published approaches in terms of distinguishing spoof attacks and speaker identification. Furthermore, the SA-SASV EER is reduced from 6.05\%, produced by previous state-of-the-art approaches, to 4.86\% when no ensemble strategy is used. The article argues that a larger set of data and distinct encoders will further improve the EER.

\section{Countermeasures}
This section provides a detailed discussion of the selected countermeasures used for comparative analysis of the voice spoofing countermeasures. For this comparative analysis, we have selected a variety of audio features, including the baseline features of the ASVspoof challenges, which are employed in the voice spoofing detection literature. However, it should be noted that the individual features are compared to one another, excluding feature fusion, to investigate the effectiveness of the features themselves and not their complementary interactions. Each feature was carefully crafted to offer as much coverage as possible while maintaining a relatively small comparative list, and each has its own distinct approach to spoofing detection. To fairly choose this list of auditory features, we conducted a survey of the literature, from 2015 to the present to determine the 14 most effective and commonly used feature descriptors. We employ the frequently used machine learning-based classifiers, i.e., GMM, SVM, and evaluate the performance of the features on the recently released deep learning CNN and CNN-GRU classifiers. The next section discusses the extraction of the SOTA countermeasures. 
\subsection{Constant Q cepstral coefficient (CQCC)) \cite{Todisco2017ConstantQC}}
The ASVspoof 2019 challenge defined two baseline systems; one used the Constant Q cepstral coefficient (CQCC) while the second employed linear frequency cepstral coefficients (LFCC) \cite{Todisco2017ConstantQC} with a GMM as the back-end classifier. CQCC features use the constant q transform (CQT) that allows the capture of variable spectral resolution, dependent on frequency, i.e., increased temporal resolution at higher frequencies and lower at lower frequencies. This variable resolution analysis allows for speech characteristics to be better captured at differing frequencies. A CQCC extraction was performed by first applying CQT to the windowed audio sample, and then a fast fourier transform (FFT) was used to obtain the power spectrum of the obtained windows, which were then log-scaled. After log scaling, uniform re-sampling was performed, and finally a discrete cosine transform (DCT) was used to obtain the final cepstral coefficients. The extraction of CQCC features was performed using the following method:\\

\begin{equation}
    X^{CQ}(k,n) = \sum_{j=n-\floor*{N_k/2}}^{n+\floor*{N_k/2}} x(j)a^{\*}_{k}(j-n+N_k/2)
\end{equation}

\begin{equation}
    a_{k}(n) = \frac{1}{C}(\frac{n}{N_{k}})exp[i(2\pi n \frac{f_{k}}{f_{s}} + \Phi_{k})]
\end{equation}

\begin{equation}
    C = \sum_{l = -\floor*{N_{k}/2}}^{N_{k}/2} w(\frac{l+N_{k}/2}{N_{k}})
\end{equation}

\begin{equation}
    f_{k} = f_{1}2^{\frac{k-1}{B}}
\end{equation}

\begin{equation}
    Q = \frac{f_{k}}{f_{k+1}-f_{k}} = (2^{1/B} - 1)^{-1}
\end{equation}

\begin{equation}
     N_{k} = \frac{f_{s}}{f_{k}}Q 
\end{equation}

\begin{equation}
    CQCC(p) = \sum_{l=1}^{L} log |X^{CQ}(l)|^{2}cos[\frac{p(l-\frac{1}{2})\pi}{L}]
\end{equation}
where  $ a^{\*}(n) $ represents a complex conjugate of $ a^{\*}(n) $, $ N_{k} $ denotes a variable window lengths, w(t) is a window function (i.e. the Hanning Function). $f_{k}$ is the central frequency, f1 refers to the center frequency of the lowest bin, and B determines the number of bins, or octaves.

\subsection{Mel-Frequency cepstral coefficients (MFCC)  \cite{muda2010voice}}
The Mel-Frequency Cepstral Coefficients (MFCC) \cite{muda2010voice} feature is well known and has been used successfully in various forms of audio recognition. To obtain MFCC features, the audio sample is divided into a number of equal windows. Following that, an FFT is performed to acquire the spectrum, which is then passed through a Mel-scaled filterbank made of triangle bandpass filters. Log-scaling is then performed, and the cepstral coefficients are obtained using a DCT.

MFCCs are computed as follows:\\
\begin{equation}
    MFCC(q) = \sum_{m=1}^{M} log[MF(m)]cos[\frac{q(m-\frac{1}{2})\pi}{M}]
\end{equation}

\begin{equation}
    MF(m) = \sum_{k=1}^{K}|X^{DFT}(k)|^{2} H_{m}(k) 
\end{equation}

where k refers to the discrete fourier transform (dft) index and H(k) denotes the triangular weighting-shaped function for the m-th Mel-scaled band-pass filter.

\subsection{Inverse Mel-Frequency cepstral coefficient (IMFCC) \cite{chakroborty2009improved}}
One variation of MFCC, the inverse Mel-Frequency cepstral coefficients \cite{chakroborty2009improved}, has also been employed for voice spoofing. These features are capable of capturing important traits of the audio signals by analyzing the high frequency components by using an inverted Mel-scaled filter bank made up of triangular band pass filters. IMFCC is obtained as follows:\\
\begin{equation}
    IMF(m) = \sum_{k=1}^{K}|X^{DFT}(k)|^{2} H_{m}(k) 
\end{equation}

\begin{equation}
    IMFCC(q) = \sum_{m=1}^{M} log[IMF(m)]cos[\frac{q(m-\frac{1}{2})\pi}{M}]
\end{equation}

where k and H(k) refer to the index and triangular weighting-shaped function for the m-th Inverted Mel-scaled band pass filter.

\subsection{ Linear Frequency Cepstral Coefficient (LFCC) \cite{zhou2011linear}}
Another common variant of MFCC is the Linear Frequency Cepstral Coefficient (LFCC) \cite{zhou2011linear}, which is derived via the same process as MFCC but uses a linear frequency filter bank in place of the Mel filter bank. In the linear frequency filter bank, the bandwidth of the triangular filter remains constant, in contrast to the Mel filter bank, which changes slope over the frequency range. Because of this, LFCC has been proven to be more robust than MFCC in detecting artifacts at higher frequencies\cite{10.1007/978-3-319-67934-1_9}.\\

\begin{equation}
    f(m) = (\frac{N}{F_s})(f_1 + m\frac{f_h-f_l}{M+1})
\end{equation}
where f(m) are the frontier points of the linear filterbank.\\

\subsection{Magnitude-based Spectral Root Cepstral Coefficients (MSRCC) and Phase-based Spectral Root Cepstral Coefficients (PSRCC) \cite{tapkir2018novel}}
Magnitude-based Spectral Root Cepstral Coefficients (MSRCC) and Phase-based Spectral Root Cepstral Coefficients (PSRCC) are spectral root features \cite{tapkir2018novel} that capture both the magnitude and phase information, respectively, to produce distinct traits in bonafide and spoofed audio. These features are computed as follows:\\

\begin{equation}
    MSRCC(q) = \sum_{m=1}^{M}(MFM(m))^{\gamma}cos[\frac{q(m-\frac{1}{2})\pi}{M}]
\end{equation}

\begin{equation}
    MFM(m) = \sum_{k=1}^{K}|S(k)|^{2}H_{m}(k)
\end{equation}

\begin{equation}
     PSRCC(q) = \sum_{m=1}^{M}(MFP(m))^{\gamma}cos[\frac{q(m-\frac{1}{2})\pi}{M}
\end{equation}

\begin{equation}
    MFP(m) = \sum_{k=1}^{K} \measuredangle{S(k)}H_{m}(k)
\end{equation}\\
where $S(k)$ is the k-point of the DFT of signal $s(n)$ and $H_{m}(k)$ is the triangular weighting-shaped function for the m-th Mel-scaled bandpass filter. \\

\subsection{Spectral Centroid Magnitude Coefficients (SCMC) and Spectral Centroid Frequency Coefficients (SCFC) \cite{kua2010investigationOS}}
Spectral Centroid Magnitude Coefficients (SCMC) and Spectral Centroid Frequency Coefficients (SCFC) are the two-dimensional representations of sub-band energy in the speech spectrum \cite{kua2010investigationOS}. These two-dimensional features include the centroid extraction of magnitude and frequency-based features from the speech spectrum. Spectral centroid features were previously employed for speaker recognition, and sub-band spectral features have demonstrated significant efficacy in noisy speech recognition. Centroid features carry formant-related information, which has been shown to be robust to the presence of noise in the vocal sample \cite{kua2010investigationOS}. Spectral centroid frequency (SCF) is the weighted average frequency for a specific sub-band, where the weights are the normalized energy of each frequency component of the sub-band. Spectral centroid magnitude (SCM) is the weighted average magnitude for a given sub-band. These coefficients are derived as follows: 

\begin{equation}
    SCM_{t}^{i} = \frac{\sum_{k=1}^{M/2+1} f(k)S_{t}(k)w_{i}(k)}{\sum_{k=1}^{M/2+1}S_{t}(k)w_{i}(k)}
\end{equation}

\begin{equation}
    SCF_{t}^{i} = \frac{\sum_{k=1}^{M/2+1} f(k)S_{t}(k)w_{i}(k)}{\sum_{k=1}^{M/2+1}f(k)w_{i}(k)}
\end{equation}
where $S_{t}(k)$ and $f(k)$ are the power spectral magnitude of the t-th frame, and normalized frequency corresponding to the k-th frequency component, respectively. SCFCs are extracted directly using linear rectangular filter banks, whereas SCM is passed through log scaling and DCT to obtain SCMC features.\\
 \begin{figure*}[t]
        \centering
        \includegraphics[width=14cm]{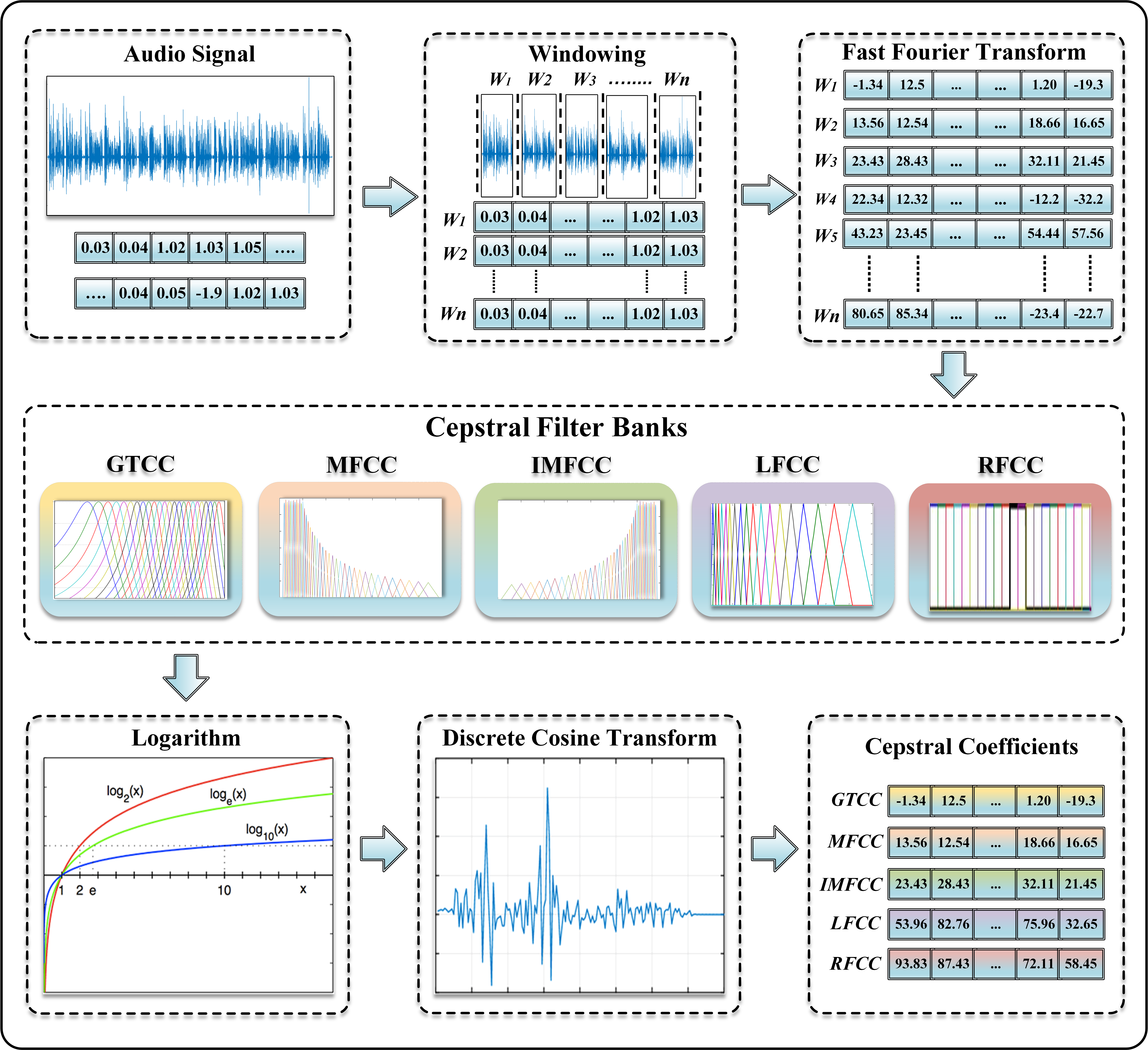}
        \caption{Extraction of GTCC, MFCC, IMFCC, LFCC and RFCC Features.}
        \label{fig:Figure5}
\end{figure*}
 \subsection{All-Pole Group Delay Function (APGDF)  \cite{rajan2013using}} 
 Unlike the features derived from the magnitude spectrum, e.g., MFCCs, the All-Pole Group Delay Function (APGDF) is extracted from the phase spectrum, which is largely ignored by the commonly used cepstral coefficients, e.g., GTCC, LPCC, and CQCC \cite{ rajan2013using}. APGDF provides an efficient way to utilize the information from the voice signal's phase spectrum. Group delay functions employ parametric all-pole models instead of spectral transforms. APGDFs enhance recognition accuracy and give comparable results to traditional magnitude-based MFCC features by integrating all-pole analysis, filters, and group delay functions. Moreover, APGDF features are computationally less complex than the recently cited MGDF features \cite{ murthy2003modified}. APGDF features are extracted as follows: 
 \begin{equation}
    APGDF(i) = \frac{A_{P}(G)D_{P}(G)+A_{I}(G)D_{I}(G)}{|A(P)|^2}
\end{equation}
\begin{equation}
    A_{P}=|A_{P}|\sigma^{i\vartheta(P)}
\end{equation}
Where $A_{P}(G)$ and $D_{P}(G)$ are Fourier transforms, $|A_{P}|$ refers to the magnitude spectrum and $i\vartheta(P)$ represents the phase spectrum. Finally, a DCT is used to convert the group delay function into cepstral coefficients \cite{ rajan2013using}.\\
\subsection{Rectangular Filter Cepstral Coefficient (RFCC)}
Another tested feature is the Rectangular Filter Cepstral Coefficient (RFCC). Similar to MFCC features, RFCCs are extracted using a rectangular window spaced on a linear scale instead of triangular filters on the mel-scale. RFCC features were initially proposed for use in automatic speaker recognition systems operating under noisy conditions.

\begin{figure*}[t]
        \centering
        \includegraphics[width=16cm]{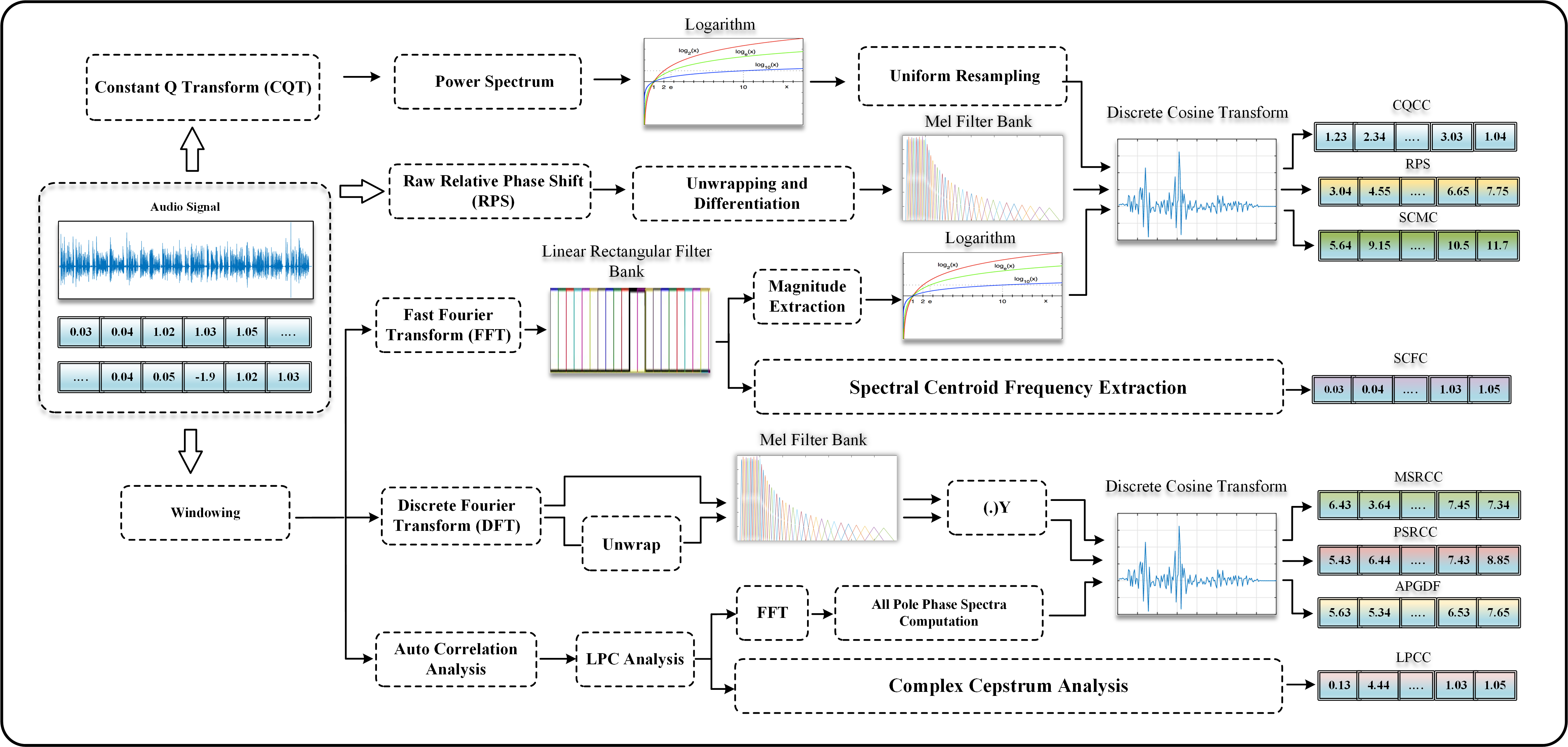}
        \caption{Extraction of CQCC, RPS, SCMC, SCFC, MSRCC, PSRCC and APGDF Features.}
        \label{fig:Figure6}
\end{figure*}
\subsection{Linear prediction cepstral coefficients (LPCC) \cite{wong2001comparison}}
Linear Prediction Cepstral Coefficient (LPCC) features have been historically used to capture the emotion information found in speech. Unlike other cepstral features, LPCCs use linear prediction analysis prior to cepstral analysis. The audio signal first undergoes band limiting, then a high-emphasis filter is applied. After high-emphasis, a 30ms hamming window is applied to the windows at a 10ms interval. The next step is to calculate the first through tenth linear predictor coefficients through the use of auto-correlation. Next, these linear predictor coefficients are transformed into the cepstral coefficients by using the following equations:
\begin{equation}
    \begin{split}
        & c_{1} = a_{1} \\ 
        & c_{n} = \sum_{k=1}^{n-1}(1-k/n)a_{k}c_{n-k}+a_{n}\ \ \ \  1 < n \le p
    \end{split}
\end{equation}
\begin{equation}
    h(t)=\sum_{k=1}^{N}A_{k}cos(\phi_{k}(t))\ \ \ \ \phi_{k}=2\pi kf_{0}t+\theta_{k}
\end{equation}
where $c_{i}$ and $a_{i}$ are the i-th cepstral and linear predictor coefficients, respectively.

\subsection{Gammatone Cepstral Coefficients (GTCC) \cite{valero2012gammatone}}
Gammatone Cepstral Coefficients (GTCCs) \cite{valero2012gammatone} are similar to MFCC features, but are much more robust to noise than MFCC, offering superior performance in classification tasks. A gammatone filter provides more frequency components in the low-frequency range with narrow bandwidth while providing fewer frequency components in the high-frequency range with wider bandwidth, effectively revealing the spectral information. GTCC features are computed by applying an FFT to each frame in order to get the spectrum. The spectra of each window are then passed through a gammatone filter bank to obtain the energy at each subband. A logarithmic function is then applied to these energies, and finally, DCT is applied to obtain the GTCCs. 13 to 20 coefficients are usually extracted and are considered optimal for audio signal analysis. GTCCs are extracted using the following equation:
\begin{equation}
    GTCC_{m} = \sqrt{\frac{2}{N}}\sum_{n=1}^{N}log(X_{n})cos[\frac{\pi n}{N}(m-\frac{1}{2})]\ \ \ \ 1 \le m\le M
\end{equation}
where $X_{n}$ is the signal energy of the nth spectral band, N is the number of gammatone filters used, and M is the number of GTCCs to be extracted.

\subsection{Relative Phase Shift (RPS) \cite{saratxagainproceedings}}
Relative Phase Shift (RPS) features \cite{saratxagainproceedings} are a representation of harmonic phase information. A harmonic analysis model represents each frame of a signal as a sum of sinusoidal harmonically related to pitch and fundamental frequency. The RPS representation calculates the difference between the instantaneous phase of every harmonic and fundamental component at a fixed point over the period of the signal. RPS features reveal a structured pattern in the phase information of voiced segments, and can be derived as follows:

\begin{equation}
    h(t)=\sum_{k=1}^{N}A_{k}cos(\phi_{k}(t))\ \ \ \ \phi_{k}=2\pi kf_{0}t+\theta_{k}
\end{equation}

A graphical representation for the extraction of features is presented in Figure \ref{fig:Figure5} and \ref{fig:Figure6}. The next section contains the experimental details, data sources and the system information used to perform the experimentation and the comparative analysis of the features.

\section{Publicly Available Datasets}
\begin{figure*}[t]
    \centering
    \includegraphics[width=13 cm]{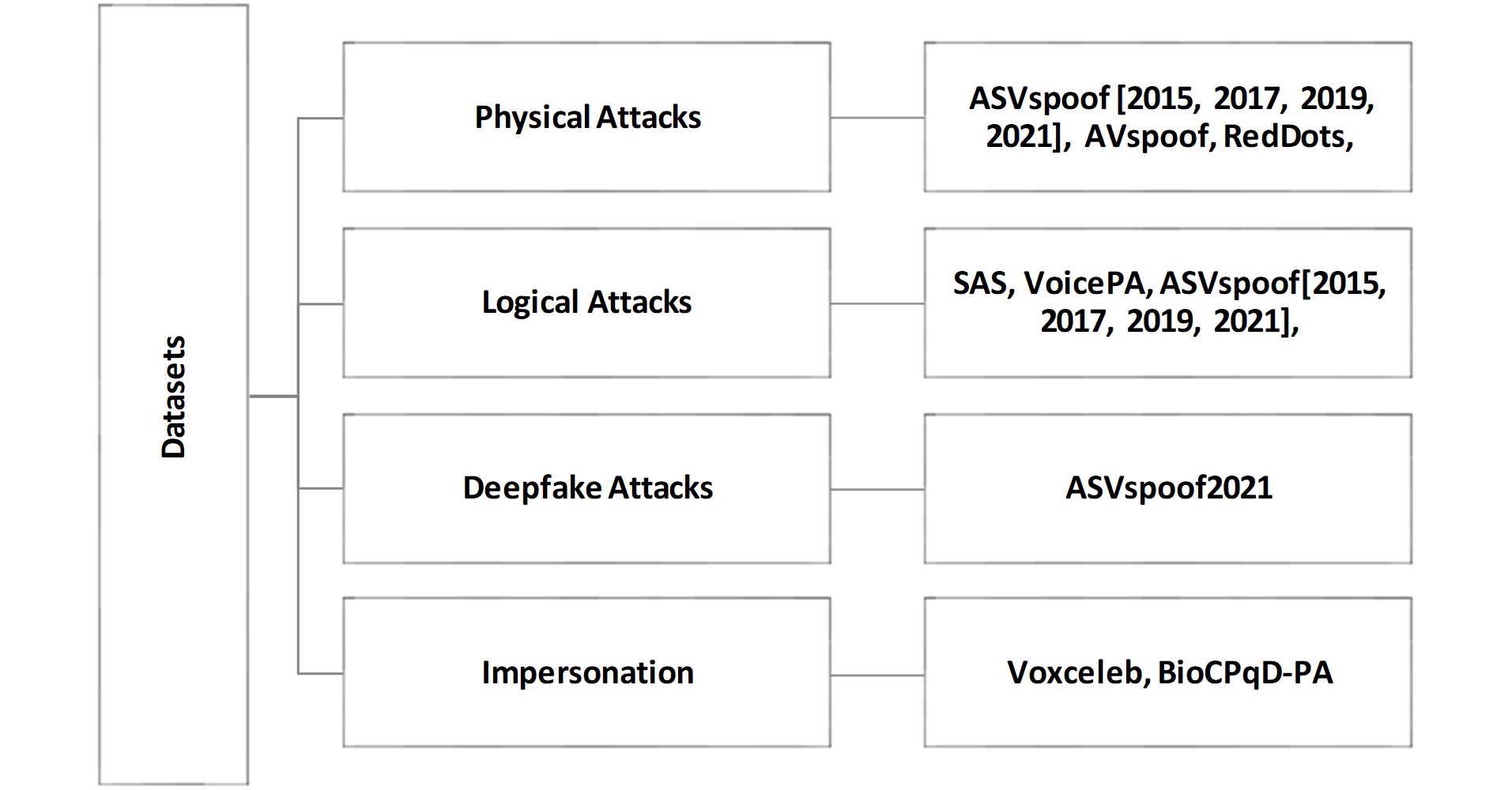}
    \caption{Existing Datasets used for Automatic Speaker Verification Systems.}
     \label{Figure7}
\end{figure*}

This section of the paper discusses and identifies the various datasets used in cutting-edge ASV systems. Early stages of voice spoofing detection research involved speech and speaker recognition databases, i.e., YOHO \cite{kreuk2018fooling}, NIST \cite{alegre2013spoofing}, and WSJ \cite{ergunay2015vulnerability}. However, to accurately account for research progress, there was a dire need for a common dataset as well as a performance metric to evaluate spoofing countermeasures. Consequently, this need was discussed and addressed at the INTERSPEECH 2013 special session on spoofing and ASV countermeasures. This special session motivated the research community to organize the first Automatic Speaker Verification Spoofing and Countermeasures Challenge, ASVspoof, in 2015, which took place at INTERSPEECH that year. The dataset released for this challenge included two types of spoofing attacks: synthetic speech (SS) and voice conversion (VC). In the years following, three additional ASVspoof challenges were organized: ASVspoof 2017, ASVspoof 2019, and ASVspoof 2021, each with publicly available datasets for download. There are several common publicly available datasets which are used by voice PAD researchers, i.e., ASVspoof, AVspoof, ReMASC, Spoofing and anti-spoofing (SAS) corpus, RedDots, Vox Celeb, voicePA, and BioCPqDPA, among others. We made a concerted effort to cover all of the existing datasets (2015–2022) used for spoof detection and countermeasure development. Details of the publicly available datasets which address spoofing attacks are presented in Figure \ref{Figure7}.

\subsection{Spoofing and Antispoofing (SAS) corpus \cite{wu2015sas}}
The Spoofing and Antispoofing (SAS) corpus \cite{wu2015sas} contains a wide range of spoofed speech samples generated using nine different approaches, two of which are speech synthesis, and the other seven are voice conversion. This database \cite{wu2015sas} contains two protocols: one for testing the ASV system and another for generating spoofed speech sounds. Consequently, they enable the speech synthesis community to create spoofing materials attentively while remaining unaware of speaker verification spoofing and anti-spoofing. Periods of silence not found in natural speech were removed from the samples, resulting in a more realistic Speech Synthesis (SS) and Voice Conversion (VC) spoof corpus (Wu et al., 2015a, 2015b, 2015c). The SAS corpus contains speech produced using various spoofing methods, represented in 300,000 samples of each type. Without using a countermeasure, ASV systems were extremely vulnerable to spoofing attacks in SAS \cite{wu2015sas}.

\subsection{RedDots \cite{Lee2015TheRD,Kinnunen2017RedDotsRA}}
The RedDots project \cite{Lee2015TheRD} was launched as a follow-up to a special session at INTERSPEECH 2014, with collaboration from multiple sites. Its goal is to collect speech data through mobile crowd sourcing, which allows for a larger population and greater diversity. On January 29, 2015, the project was launched. At the time of our investigation, the project had 89 speakers from 21 countries, 72 men and 17 women, for a total of 875 complete sessions. The purpose of this special session is to bring together research efforts to investigate potential approaches and gain a better understanding of speaker/channel phonetic variability. The RedDots dataset \cite{Lee2015TheRD} is intended to have a higher number of recordings per session, with shorter sessions spent on each. One goal is to collect 52 sessions per speaker, one session per week, for a year. Because of this, each session is limited to two minutes, with a total of 24 sentences (10 common, 10 unique, 2 free choices, and 2 free texts) for each session. This dataset \cite{Lee2015TheRD} includes a diverse set of inter- and intra-speaker types. Following that, the replayed RedDots database \cite{Kinnunen2017RedDotsRA} was created by re-recording the original corpus utterances under different environmental conditions. Both of these databases assist in the development of replay-resistant ASV systems because the original RedDots provided genuine utterances. An overview of the RedDots database \cite{Lee2015TheRD} is presented in Table \ref{tab:All4datasets}.

\subsection{AVspoof \cite{korshunov2016overview}}
The AVspoof dataset is designed to assist ASV systems in the development of anti-spoofing techniques. This database was used in the BTAS 2016 \cite{korshunov2016overview,ergunay2015vulnerability} challenge. The AVspoof database includes replay spoofing attacks in addition to synthetic speech and VC spoofing attacks. The replay attacks are generated by various recording devices. The SS attacks were generated by the following techniques, i.e., Hidden Markov Model (HMM) and Festvox, which account for the vast majority of the VC attacks. These sessions are recorded by participants in a variety of environments and with a variety of recording devices. Speakers are instructed to read out sentences, phrases, and speak freely about any topic for 3 to 10 minutes. To make competition more difficult, "unknown" attacks are included in the test set \cite{korshunov2016overview}. The organizers of the challenge provide a baseline evaluation system based on the open source Bob toolbox \cite{korshunov2016overview}. In the baseline system, simple spectrogram-based ratios serve as features, and logistic regression is used as a pattern classifier \cite{korshunov2016overview}. The statistics of the database are summarized in Table \ref{tab:AVspoof}.

\begin{table}[H]
\caption{A summary of AVspoof Database.}
\label{tab:AVspoof}
\centering
\begin{tabular}{p{2.2cm} | p{1.3cm} | p{1.3cm} | p{1.3cm}}
  \hline
 Subset & Genuine  & Spoof (PA) & Spoof (LA) \\
 \hline
 Training & 4973 & 38580 & 17890\\
 Development & 4995 & 38580 & 17890\\
 Evaluation & 5576 & 43320 & 20060 \\
 \hline
 \end{tabular}
 \end{table}

\subsection{VoxCeleb \cite{nagrani2017voxceleb}}
VoxCeleb \cite{nagrani2017voxceleb} is an audio-visual dataset comprised of short clips of human voices extracted from YouTube interview videos. Each segment lasts at least 3 seconds. VoxCeleb features speech from people of various ethnicities, accents, professional backgrounds, and ages. 61\% of speakers are male and 39\% are female. The data was collected randomly, with ambient noise, laughter, overlapping speech, pose deviation, and a variety of lighting conditions. This dataset is available in two versions: VoxCeleb1 \cite{nagrani2017voxceleb} and VoxCeleb2 \cite{chung2018voxceleb2}. Each has audio files, face clips, metadata about speakers, and so on, in the training and testing sets. The Finnish-language sets assist ASV systems in the detection of mimicry attacks \cite{vestman2020voice}. The details of the dataset \cite{nagrani2017voxceleb} are shown in table \ref{tab:All4datasets}.

\subsection{voicePA \cite{korshunov2018use}}
The voicePA dataset was created with the assistance of the AVspoof dataset. Its bonafide data is a subset of the AVspoof dataset's genuine data, uttered by 44 speakers in four recording sessions held in various settings \cite{wang2020asvspoof}. These sessions were recorded using high-quality microphones from a laptop, a Samsung S3, and an Apple smartphone 3GS. Spoofed data consists of 24 different types of presentation attacks that are captured using five devices in three different environments. These spoof utterances are based on real-world data. The original dataset's SS and VC spoofed audio samples are also replayed in the voicePA dataset \cite{korshunov2018use}. Table \ref{tab:All4datasets} contains detailed information about the voicePA \cite{korshunov2018use} dataset.

\subsection{BioCPqD‑PA \cite{korshunov2018use}}
The Portuguese language BioCPqD‑PA dataset \cite{korshunov2018use} was collected by recording 222 people in a variety of environmental conditions. The dataset was comprised of 27,253 authentic recordings and 391,687 samples, which had been subjected to a presentation attack. In the presentation audio data, one laptop was used with 24 different setups consisting of 8 loudspeakers and 3 microphones, while another single laptop was used to capture real data. This dataset \cite{korshunov2018use} was divided into three parts: training, development, and evaluation. Each part was collected using a different set of microphones and loudspeakers, and each set was comprised of the audio recording from all of the participating speakers. An overview of the BioCPqD‑PA \cite{korshunov2018use} dataset is presented in Table \ref{tab:All4datasets}. 

\begin{table}[H]
     \caption{A summary of the existing databases.}
     \label{tab:All4datasets}
\centering
\begin{tabular}{p{1.6cm}|p{1cm}|p{1.2cm}|p{1.5cm}|p{1.3cm}}
  \hline
Dataset  & Language & Attacks & Speakers & Utterances \\
  \hline
 RedDots & 5  & Replay & 3750 & 5000 \\
 VoxCeleb & 6+ & Mimicry & 1251 (V1), 7000+ (V2) & 100,000 (V1), 1,000,000 (V2) \\ 
 VoicePA & - & SS, VC & 44 & - \\ 
  BioCPqD-PA & 1 & 222 & 3750 & 418,940 \\\hline
     \end{tabular}
 \end{table}
 
 \begin{table}[H]
\caption{A summary of ASVspoof Challenge 2015 database.}
\label{tab:Asvspoof2015}
\centering
\begin{tabular}{p{1.6cm}|p{1cm}|p{1cm}|p{1cm}|p{1cm}}
  \hline
  & \multicolumn{2}{c|}{Speaker} & \multicolumn{2}{c}{Utterances}\\
  \hline
 Subset & Male  & Female & Genuine & Spoofed \\
 \hline
 Training & 10 & 15 & 3750 & 12625 \\
 Development & 15 & 20 & 3497 & 49875\\ 
  Evaluation & 20 & 26 & 9404 & 193404 \\ \hline
     \end{tabular}
 \end{table}

\subsection{ASVspoof 2015 \cite{wu2015asvspoof} }
The ASVspoof 2015 Challenge database is the first significant release for research into spoofing and countermeasures \cite{wu2015asvspoof}. For LA attacks, the database contains natural and spoofed speech generated by speech synthesis and VC. There are no noticeable effects from the channel or background noise. The database is split into three sections: training, development, and evaluation. The evaluation subset is made up of both known and unknown attacks. The known attacks contain the very same five algorithms that were used to generate the development dataset and are thus referred to as known (S1-S5) attacks. Other spoofing algorithms are also included in unknown (S6-S10) attacks that are directly used in the test data. Table \ref{tab:Asvspoof2015} contains a detailed description of the challenge database.

\subsection{ASVspoof 2017 \cite{kinnunen2017asvspoof}}
The ASVspoof 2017 Challenge dataset \cite{kinnunen2017asvspoof} was constructed on the RedDots dataset \cite{Lee2015TheRD}. The dataset \cite{kinnunen2017asvspoof} contains replayed data samples with text-dependent speech. It included the voices of 42 different speakers, recorded using 61 combinations of distinct recording devices, replay devices, and environmental conditions. It was collected over the course of 179 sessions. The original ASVspoof 2017 database \cite{kinnunen2017asvspoof} contained some inconsistencies, but the issue was resolved in ASVspoof 2017 Version 2.0 \cite{delgado2018asvspoof}. In addition to the corrected data, a more detailed description of recording and playback devices, as well as acoustic environments, was provided \cite{delgado2018asvspoof}. The details of the {ASVspoof 2017 \cite{kinnunen2017asvspoof}} dataset are presented in Table \ref{tab:Asvspoof2017}. 

\begin{table}[H]
\caption{A summary of ASVspoof Challenge 2017 database version 2.0}
\label{tab:Asvspoof2017}
\centering
\begin{tabular}{ p{1.9cm} | p{1.5cm} | p{1.5cm} | p{1.5cm}}
  \hline
 Subset & Speaker  & Genuine & Spoofed \\
 \hline
 Training & 10 & 1507 & 1507\\
 Development & 8 & 760 & 950\\
 Evaluation & 24 & 1298 & 12008 \\
 \hline\end{tabular}
 \end{table}
 
\subsection{ReMASC \cite{gong2019remasc}}
The ReMASC (Realistic Replay Attack Microphone Array Speech Corpus) is a database of speech recordings developed to support research into and the security of voice-controlled systems. The ReMASC database is comprised of authentic and replayed recordings of speech samples, captured in actual circumstances, and utilizes cutting-edge voice assistant development kits. In particular, the ReMASC dataset contains recordings from four systems, each with a different transmitter and receiver, under a range of atmospheric situations, with varying levels of background noise and relative speaker-to-device locations. This is the first database that was specifically developed to safeguard voice-controlled systems (VCS) from various types of replay attacks in varied contexts. Table \ref{tab:ReMASC} contains the sampling details of the eMASC \cite{gong2019remasc} dataset.
\begin{table}[H]
\caption{A summary of ReMASC Database.}
\label{tab:ReMASC}
\centering
\begin{tabular}{p{2.0cm} | p{1.5cm} | p{1.5cm} | p{1.8cm}}
  \hline
 Environment & Subject  & Genuine & Replayed \\
 \hline
 Outdoor & 12 & 960 & 6900\\
 Indoor1 & 23 & 2760 & 23104\\
 Indoor2 & 10 & 1600 & 7824 \\
 Vehicle & 10 & 3920 & 7644\\
 Total & 55 & 9240 & 45472 \\
 \hline      
\end{tabular}
 \end{table}

\subsection{ASVspoof 2019 \cite{wang2020asvspoof}}
The ASVspoof 2019 challenge \cite{wang2020asvspoof} is an extension of the previously held ASVspoof challenges and focuses on countermeasures for all three major attack types, namely, SS, VC, and replay. It is divided into the logical access (LA) and physical access (PA) subsets. LA contains TTS and VC spoof speech samples, and PA has replay-spoof speech. Both of these subsets are further partitioned into training, development, and evaluation subsets. The training subset is generated by 8 males and 12 females; the development subset by 4 males and 6 females, and the evaluation subset by 21 males and 27 females. The spoofed speech signals are generated using one of the two VC and four speech synthesis algorithms. Details of the ASVspoof 2019 \cite{wang2020asvspoof} dataset are shown in Table \ref{tab:Asvspoof2019}.

\subsection{ASVspoof 2021 \cite{delgado2021asvspoof}}
ASVspoof 2021 \cite{delgado2021asvspoof} is the fourth offering in the series of spoofing challenges designed to promote spoofing research and the development of countermeasures to protect automatic speaker verification systems from manipulation. ASVspoof 2021 \cite{delgado2021asvspoof} introduces a new task involving deepfake speech detection, in addition to a continued focus on logical and physical access tasks, and has a number of advancements over previous editions. ASVspoof 2021, in particular, is divided into three sub-challenges. The first is a logical access sub-challenge that builds on the 2019 challenge by emphasising robustness to channel variation; the second is a physical access sub-challenge, similar to the 2019 setup, but with recordings made in real-world physical environments. The third is a speech deepfake detection sub-challenge (no ASV). ASVspoof 2021 contains technically difficult data to encourage broad generalization of countermeasures.

\begin{table}[H]
\caption{A summary of ASVspoof Challenge 2019 database.}
\label{tab:Asvspoof2019}
\begin{tabular}{p{1.5cm}|p{0.5cm}|p{0.7cm}|p{0.8cm}|p{0.8cm}|p{0.8cm}|p{0.8cm}}
  \hline
  & \multicolumn{2}{l|}{Speaker} & \multicolumn{2}{l|}{LA Attacks}  & \multicolumn{2}{l}{PA Attacks}  \\  
  \hline
 Subset & Male  & Female & Genuine & Spoofed & Genuine & Spoofed \\
 \hline
 Training & 8 & 12 & 2580 & 22800 & 5400 & 48600\\
 Development & 8 & 12 & 2548 & 22296 & 5400 & 24300\\ \hline
 Evaluation & - & - & \multicolumn{2}{l|}{71747} & \multicolumn{2}{l}{137457}  \\
 \hline
     \end{tabular}

 \end{table}

The logical access (LA) task is similar to that of ASVspoof 2017 and 2019, but this version includes the computation and transmission of text-to-speech (TTS) and voice conversion (VC) attacks. In comparison to ASVspoof 2017, the physical access (PA) task includes genuine and replayed samples but with a more tightly controlled setup. The new speech deepfake task (DF), similar to the LA task, includes compressed audio. The protocols used in the training and development sections are the same as they were in ASVspoof 2019, and therefore ASVSpoof 2021 does not include development or training subsets. There are new metrics for the evaluation partition, including a slightly revised t-DCF metric for the LA and PA tasks. Whereas the EER is used for the evaluation of DF tasks.

\subsection{Voice Spoofing Detection Corpus (VSDC) \cite{baumann2021voice}} 
The VSDC dataset \cite{baumann2021voice} was designed to be a standard dataset comprising both real audio recordings from various surroundings and microphones, as well as spoofed audio files generated through various controlled environments, spoofing realistic scenarios. This data can be analyzed and utilized to develop countermeasures to avoid audio spoofing attacks in the replay category. Since an Internet of Things (IoT) device may be exploited as a point of replay to another device, audio replay attacks have become more significant. This offers a perfect atmosphere for carrying out repeat attacks. This data set was generated with the intention of simulating these attacks in a controlled environment. While the primary goal of this dataset is to detect multi-hop replay attacks, it is not restricted to that. It could be used to investigate typical replay attacks, the influence of different microphones and surroundings on an audio file, or how an individual's vocal range affects the accuracy of a voice control system. The details of the number of sample and development architecture can be found in \cite{baumann2021voice}.

\section{Performance Evaluation Parameters}
This section describes the performance evaluation criteria used in this survey, and the specific metrics are defined in the following subsections. After an extensive review of published work, we observed that Equal Error Rate (EER) was the primary criterion used for the performance evaluation of voice spoofing and ASV systems. Of the articles surveyed for this paper, more than three-quarters evaluated their presented work using EER, for instance \cite{10.1007/978-3-030-43575-2_9}. A small number of the investigated papers picked accuracy as the sole performance rating criterion, however, the majority of recent work uses multiple rating criteria to evaluate the performance of the presented method, e.g., EER with min-tDCF, EER with Half Total Error Rate (HTER), and False Match Rate (FMR) with False Non-match Rate (FNMR).  As not all research works were examined using the same criteria, comparative analysis of the results became problematic. 
\subsection{Equal error rate (EER)}
An ASV system either approves or rejects the specified identities. A categorization might be correct or erroneous in four ways. These are True Acceptance (TA), True Rejection (TR), False Acceptance (FA), and False Rejection (FR). TA and TR are favorable outcomes, but FR and FA are damaging to the system. These possibilities are determined by a preset threshold \cite{Todisco2017ConstantQC}. In the case of FA, a faked speech sample with a score higher than or equal to the preset threshold is allowed. Conversely, in the case of FR, a speech sample with a score less than the predefined threshold is rejected. To evaluate ASV performance, the Equal Error Rate (EER), the value at which the False Acceptance Rate (FAR) and False Rejection Rate (FRR) become equal, is employed.
\begin{equation}{
    FAR=\frac{FA}{TA}\\}
\end{equation}
\begin{equation}{
    FRR=\frac{FR}{TT}\\}
\end{equation}
  
where FAR and FRR denote the false acceptance and false rejection rate, FA and TA represent the number of false acceptance and false rejection, and TA and TT refer to the total number of false and genuine speech samples.
\subsection{Detection Error Tradeoff (DET) curve}
A Detection Error Tradeoff (DET) graph is a graphical representation of errors for classification model systems that compares the false rejection and the false acceptance rate. The x- and y-axes are scaled non-linearly by their standard normal deviations (or simply by logarithmic transformation), resulting in tradeoff curves that are more linear than ROC curves and use the majority of the area to highlight substantial differences in the crucial operating zone. With its effectiveness for binary classification, ASV systems are initially evaluated using the DET curve. It generates EER curves by plotting FAR on the x-axis and FRR on the y-axis.

\subsection{Half Total Error Rate (HTER)}
Spoofed speech with a higher score than the predefined threshold will be misidentified as bonafide samples, whereas bonafide samples with a lower score will be incorrectly classified as spoofed speech samples. Since the two errors, FAR and FRR, are inversely connected, it is essential to depict performance as a function of the threshold. The Half Total Error Rate is one such metric (HTER). HTER's computation is shown below: 
\begin{equation}{
    HTER=\frac{FAR+FRR}{2}\\}
\end{equation}

\subsection{Tandem Detection Cost Function (t‑DCF)}
The ASVspoof challenge series was created to lead anti-spoofing research for automated speaker verification (ASV). The 2015 and 2017 challenge editions featured evaluating spoofing countermeasures (CMs) in isolation from ASV using an equal error rate (EER) metric. While this was a strategic method of evaluation at the time, it had certain flaws. First, when ASV and CMs are coupled, the CM EER is not always a trustworthy predictor of performance. Second, the EER operating point is inappropriate for user authentication systems with a high target user prior but a low spoofing attack prior, such as telephone banking. As a consequence, the community will shift from CM-centric to ASV-centric evaluation using a new tandem detection cost function (t-DCF) measure. It extends the typical DCF used in ASV research to spoofing attack circumstances. The t-DCF metric is made up of six components in two parts: (Part i) false alarm(1-2) and miss costs(3-4) for both systems; and (Part ii) prior probability of target(5) and spoof trials(6) (with an implied third, non-target prior). The results of t-DCF in \cite{kinnunen2018t} justify the inclusion of the DCF-based metric in the roadmap for future ASVspoof challenges, as well as additional biometric anti-spoofing evaluations. 

\section{Comparative analysis of voice spoofing countermeasures} 
In this section of the paper, we discuss the datasets and classifiers used to differentiate between spoofed and real speech samples. The dataset subsection contains information on speech samples and corpus details that were utilized in training, development, and testing, as well as cross-corpus evaluation of countermeasures. The subsection on feature and classifier configuration describes the back end classification approach used to evaluate the effectiveness of the countermeasures. Next, the experimental analysis's environmental setup is described. 

\subsection{Dataset}
For the comparative analysis, we used two datasets to perform the experimentation: VSDC \cite{baumann2021voice} and ASVspoof 2019 \cite{wang2020asvspoof}. The VSDC dataset consists of a wide variety of microphones and speakers. Although the VSDC dataset contains types of bonafide, first-order, and second-order replay recordings, it is designed to provide a diverse set of recording and playback environments. Specifically, the microphones range from professional grade to cellphones, and the speakers for replay range from professional to local residents. In the VSDC, a total of fifteen (15) speakers, ages 18–60 years old, participated in data collection. Out of fifteen (15) speakers, ten (10) are male, and five (5) are female. Some of the speakers are not native English speakers. Each speaker recorded the original file by repeating a given set of phrases typical of commands given to VCDs. Some of the volunteers recorded these original phrase sets multiple times using different microphones in diverse environments. In total, 198 different 0PR source sets, consisting of at least 9 phrases, were recorded, resulting in 1,926 0PR source phrases being spoken.1PR and 2PR samples were obtained from 0PR samples. There were 22 different microphones used, with 14 different microphone-speaker configurations for 1PR and 8 different configurations used for 2PR. 1PR and 2PR each include 4,923 samples, giving VSDC a total of 11,772 samples. 

In contrast, ASVspoof 2019 is the larger dataset, consisting of 107 speakers. The ASVspoof 2019 dataset includes bonafide and spoof recordings, with LA and PA attacks. The recording environments are more uniform and less diverse than in VSDC, consisting of only hemi-anechoic chambers of varying room sizes. The ASVspoof2019 dataset brings additional diversity in the length of the audio recordings, both bonafide and spoofed, that VSDC does not. The ASVspoof 2019's PA dataset consists of 54,000 samples in the training data set, 29,700 samples in the validation set, and 137,457 samples in the evaluation set. The samples were recorded from 20 individuals, consisting of 12 female and 8 male participants. The recordings utilized a 16-kHz sampling rate and 16-bit resolution, along with having 27 different acoustic configurations. The configurations had variations using 3 room sizes, 3 different levels of reverberation, and 3 different speaker-to-ASV distances. For the replay sample of the data set, 9 different configurations were used, each having 3 different attacker-to-talker distances and 3 different speaker quantities. It should also be noted that configurations for the testing data set differed from those of the training and validation sets. The more detailed description of the dataset is provided in section VI and subsection J. 

\begin{table}[t]
\caption{Configuration details of the comparative countermeasures.}
     \label{tab:features_configuration}
\centering
\begin{tabular}{*{3}{c}}
 \hline
 \multicolumn{3}{c}{Features} \\
  \hline
 Feature & Type & Configuration\\
 \hline
 CQCC\cite{Todisco2017ConstantQC} & Cepstral & No. of FFT bins (per octave)=96 \\
 \hline
 LFCC \cite{zhou2011linear} & Cepstral & No. of Filters=20 \\
 \hline
 MFCC \cite{muda2010voice}& Cepstral & No. of Filters=14 \\
 \hline
 IMFCC \cite{chakroborty2009improved} & Cepstral & No. of Filters=20 \\
 \hline
 SSFC \cite{tapkir2018novel}& Cepstral & No. of Coeffs.=20  \\
 \hline
 RFCC \cite{saratxagainproceedings}& Cepstral & No. of Filters=20 \\
 \hline
  PSRCC \cite{tapkir2018novel}& Cepstral & No. of Coeffs.=13  \\
 \hline
 MSRCC \cite{tapkir2018novel}& Cepstral & No. of Filters=13 \\
 \hline
LPCC \cite{wong2001comparison}& Cepstral & LP order=20 \\
\hline
SCMC \cite{kua2010investigationOS}& magnitude & No. of Sub-bands=20\\
\hline
SCFC \cite{kua2010investigationOS}& frequency & No. of Sub-bands=20 \\
\hline
GTCC \cite{valero2012gammatone}& Cepstral & No. of Coeffs.=14 \\
\hline
RPS \cite{saratxagainproceedings} & Cepstral & No. of Coeffs.=20 \\
\hline
APGDF  \cite{rajan2013using}& Phase based & LP Order=20 \\
\hline
 \end{tabular}
\end{table}

\subsection{Feature and classifier Configuration}
To test the effectiveness of the state-of-the-art (SOTA) spoofing countermeasures, four different classifiers (two machine learning and two deep learning) comprised of a Gaussian Mixture Model (GMM) \cite{todisco2019asvspoof}, a Support Vector Machine (SVM-rbf) \cite{9107388}, a Convolutional Neural Network (CNN) \cite{jung2019replay}, and a CNN-Gated Recurrent Unit (CNN-GRU) \cite{jung2019replay} were used. The baseline GMM model given by ASVspoof2019 for CQCC and LFCC features was expanded to compute the performance of all frontend features. The GMM was trained for ten iterations, and the score of the speech samples was computed using a log-likelihood ratio. EER and min-TDCF were used to assess the effectiveness of each countermeasure with GMM baseline classification. Next, an SVM classifier with an RBF kernel was used to evaluate the performance of the SOTA countermeasures. The GMM and SVM classifiers were chosen because they had the best reported results for the respective datasets. For example, the SVM-rbf classifier \cite{9107388} achieved the best performance on the VSDC dataset, whereas CQCC-GMM and LFCC-GMM were the baseline classifiers for the ASVspoof2019 competition. In addition, CNN and CNN-GRU \cite{jung2019replay} were used to determine the performance of the SOTA countermeasures on deep learning-based classifiers. The effectiveness of each countermeasure is described below in the experiment section. Section V also completely describes the extraction of the countermeasure chosen for comparative study. Although we assured the codes' availability for reproducibility, the two-dimensional characteristics of the features was converted to one-dimensional by employing the mean average of the retrieved features. The features are detailed in the code and the ReadMe.txt file. The code and ReadMe.txt file contain details of the features. Further details of the extraction mechanism used for the frontend features are mentioned in Table \ref{tab:features_configuration}.
\subsection{Experimental Setup}
The setup for the experiments consisted of a pipeline of feature extraction and then evaluation of the extracted features on the ML and DL-based classifiers. The feature extraction and classifiers were run on Oakland University's Matilda High Performance Cluster (HPC). The standard compute nodes were used for feature extraction and model training, each of which consists of 192 GB of RAM and 40 CPU cores at 2.50 GHz. The models (GMM, SVM, CNN, and CNN-GRU) were trained and tested utilizing the HPC's GPU nodes, made up of four NVIDIA Tesla V100 16G GPUs, 192 GB of RAM, and 48 CPU cores running at 2.10 GHz. The VLFEAT@matlab API was used for the GMMs, which consisted of 512 Gaussian clusters, and the GMMs were trained for 10 iterations.

\section{Experimental Results}
In this section, we report the results of the experiments performed to determine the effectiveness of the SOTA countermeasures. The effectiveness of each countermeasure was tested against standalone as well as cross-corpus evaluation. Standalone performance analysis implies the countermeasure's capability of detecting the spoofing artifacts extant in the training samples and similar ones existing in the testing samples. In contrast, cross-corpus evaluation determines the performance of each countermeasure against unknown and distinctively configured spoofing artifacts. For a fair comparison of the countermeasures, we test their effectiveness against four different ML and DL models (GMM, SVM, CNN, and CNN-GRU) and two diverse datasets, VSDC and ASVspoof2019. The VSDC dataset is used to evaluate the performance of countermeasures on smart home-based ASV systems, where microphonic discrepancies are more frequent, while the countermeasures are also tested against the ASVspoof19 industry standard spoofing discrepancies. Lastly, we comprehensively discuss the overall performance, both standalone and cross-corpus,  of the countermeasures on all datasets and classifiers.

\subsection{Comparative Analysis of SOTA countermeasures on GMM based classifier}
In this experiment, we evaluate the performance of the SOTA countermeasure with the baseline GMM classifier provided by the ASVspoof \cite{wu2017asvspoof}  challenge community. Although the classifier is initially provided with LFCC and CQCC features, we expanded it to evaluate the SOTA countermeasures. The experimental results show that each countermeasure performs significantly differently for speech synthesis (SS), VC, and replay attacks. The results demonstrate that the samples with replay attacks are much more difficult to detect in comparison with the SS and VC samples. During the evaluation, it is observed that the performance of the SOTA countermeasures declines drastically in the presence of replay artifacts in the speech samples.

The countermeasures were considered and analyzed for LA and PA attacks in the ASVspoof19 dataset examination. The countermeasures were tested against the development and evaluation speech samples, and the results are reported in Table XIII. The results demonstrated that each countermeasure achieved a good EER and min-TDCF against the development set of each spoofing attack (LA and PA). In particular, the SCMC \cite{kua2010investigationOS} countermeasure achieved a lower EER of 0.01, while the lowest minimum t-DCF was obtained by the IMFCC \cite{chakroborty2009improved} and APGDF \cite{rajan2013using} countermeasures in the testing of development samples. However, in evaluation set testing, the APGDF \cite{rajan2013using} countermeasure achieved a lower EER and m-TDCF of 5.75 and 0.14, respectively. The results demonstrated that the APGDF \cite{rajan2013using} and CQCC \cite{Todisco2017ConstantQC} countermeasures outperformed the comparatives with an overall lower EER and min-TDCF in the development and evaluation testing \ref{tab:gmmASV}. In addition, some countermeasures performed optimally against the LA attacks comprised of SS and VC but failed to perform optimally when replay artifacts existed in the speech sample. For instance, the SCMC \cite{kua2010investigationOS} countermeasures obtained the best EER of 0.01 and the second best EER of 5.91 for the development and evaluation test sets, however, SCMC failed to provide optimal performance in the case of PA attack testing. Similarly, the EER of APGDF \cite{rajan2013using} countermeasure was higher than SCMC in the development set, but APGDF \cite{rajan2013using} achieved the overall best results in the rest of the testing. The detailed results for SOTA countermeasures are reported in Table XIII.

\begin{table}[t]
\caption{Experimental performance of the countermeasures with ASVspoof2019 and GMM based classifier.}
\label{tab:gmmASV}
\centering
\begin{tabular}{*{6}{l}}
\hline
 \multirow{2}{1cm}{Feature} & \multirow{2}{0.5cm}{Dataset}  & \multicolumn{2}{c}{Development} & \multicolumn{2}{c}{Evaluation} \\ 
                        &   & EER & m-tdcf & EER & m-tdcf  \\
 \hline
 \multirow{2}{1.5cm}{LFCC \cite{zhou2011linear}}  & ASVspoof-LA &  2.70 & 0.06 & 8.08 & 0.21  \\
                        & ASVspoof-PA & 11.9 & 0.25 & 13.5 & 0.30 \\
 \hline
 \multirow{2}{1.5cm}{CQCC \cite{Todisco2017ConstantQC}} & ASVspoof-LA & 0.43 & 0.02 & 9.57 & 0.23  \\
                          & ASVspoof-PA & 4.87 & 0.19 & 11.0  & 0.24  \\
 \hline
 \multirow{2}{1.5cm}{LPCC \cite{wong2001comparison}} & ASVspoof-LA & 2.31 & 0.06 &  10.17 & 0.28 \\
                        & ASVspoof-PA & 38.5 & 0.83 & 46.6 & 0.98  \\
 \hline
 \multirow{2}{1.6cm}{MSRCC \cite{tapkir2018novel}}& ASVspoof-LA & 9.45 & 0.19 &  10.9 &  0.28 \\
                        & ASVspoof-PA & 12.7 & 0.28 &  15.7 &  0.38 \\
  \hline
 \multirow{2}{1.5cm}{PSRCC \cite{tapkir2018novel}}& ASVspoof-LA & 9.13 & 0.14 &  10.7 &  0.21 \\
                        & ASVspoof-PA & 12.7 & 0.28&  15.7 &  0.38 \\
 \hline
 \multirow{2}{1.5cm}{SCFC \cite{kua2010investigationOS}}& ASVspoof-LA & 14.7 & 0.39 &  20.6 & 0.54 \\
                        & ASVspoof-PA & 17.6 & 0.35 &  21.6 &   0.47\\
 \hline
 \multirow{2}{1.5cm}{SCMC \cite{kua2010investigationOS}}& ASVspoof-LA & 0.01 & 0.54 & 5.91  & 0.15  \\
                        & ASVspoof-PA & 12.5 & 0.27 &  13.9 &  0.33 \\
 \hline
 \multirow{2}{1.5cm}{MFCC \cite{muda2010voice}}& ASVspoof-LA & 7.06 & 0.16 &  10.56 & 0.25  \\
                        & ASVspoof-PA & 11.5 & 0.23 &  13.7 &  0.32 \\
 \hline
 \multirow{2}{2cm}{IMFCC \cite{chakroborty2009improved}}& ASVspoof-LA & 0.04 & 0.01  & 10.9  &  0.24 \\
                        & ASVspoof-PA & 12.8 & 0.29 &  13.7 &  0.32 \\
 \hline
 \multirow{2}{1.5cm}{RFCC \cite{saratxagainproceedings}}& ASVspoof-LA & 2.71 & 0.07 &  8.06 &  0.22 \\
                        & ASVspoof-PA & 11.8 & 0.26 & 13.9   &  0.33 \\
 \hline
 \multirow{2}{1.5cm}{RPS \cite{saratxagainproceedings}} & ASVspoof-LA & 9.28 & 0.19 &  11.9 &  0.29 \\
                        & ASVspoof-PA & 15.3 & 0.32 & 14.0  &  0.36 \\
 \hline
 \multirow{2}{1.5cm}{SSFC \cite{tapkir2018novel}}& ASVspoof-LA & 8.10 & 0.16 &  10.3 &  0.27 \\
                        & ASVspoof-PA & 13.8 & 0.29 & 13.9  &  0.31 \\
 \hline
 \multirow{2}{1.5cm}{GTCC \cite{valero2012gammatone}}& ASVspoof-LA & 9.25 & 0.17 &  10.8 &  0.24 \\
                        & ASVspoof-PA & 12.7 & 0.28 &  15.7 &  0.38 \\
 \hline
\multirow{2}{1.5cm}{APGDF \cite{rajan2013using}}& ASVspoof-LA &  0.22 & 0.01 & 5.75  & 0.14  \\
                        & ASVspoof-PA & 8.66 & 0.17 & 10.6  &  0.25  \\
 \hline
\end{tabular}

\end{table}
\begin{table*}[t]
\caption{Experimental performance of the countermeasures with VSDC and GMM based classifier.}
\label{tab:GMMvsdc}
\centering
\begin{tabular}{*{8}{c}}
\hline
Features & APGDF \cite{rajan2013using} & IMFCC \cite{chakroborty2009improved} & GTCC \cite{valero2012gammatone} & CQCC \cite{Todisco2017ConstantQC}& LFCC \cite{zhou2011linear}& RPS \cite{saratxagainproceedings} & PSRCC \cite{tapkir2018novel}\\ \hline
EER & 40.29 & 50.00 &  56.00 & 20.0 & 36.33 & 55.21 & 55.11  \\ \hline
Features  & LPCC \cite{wong2001comparison} & MFCC \cite{muda2010voice} & RFCC \cite{saratxagainproceedings} & SCFC \cite{kua2010investigationOS} & SCMC \cite{kua2010investigationOS} & SSFC \cite{tapkir2018novel}& MSRCC \cite{tapkir2018novel}\\  \hline
 EER & 5.86 & 49.40 & 48.87 & 45.49 & 49.91 & 53.11 & 54.73\\
 \hline
\end{tabular}
\end{table*}

Along with the ASVspoof2019, we also examined the performance of the SOTA countermeasure with the VSDC dataset, and the results are shown in Table \ref{tab:GMMvsdc}. In the instance of the VSDC dataset, we computed the EER to assess the effectiveness of the countermeasures. The results showed that the LPCC \cite{wong2001comparison} countermeasure exhibited the best and CQCC \cite{Todisco2017ConstantQC} obtained the second best EER of 5.86 and 20.0 in the VSDC dataset testing, respectively. Although the VSDC dataset contains 1PR and 2PR replay samples, this experiment only included 1PR testing because the goal was to distinguish between spoofed and genuine samples.It was observed from the results that GTCC \cite{valero2012gammatone} features proved to be an expensive countermeasure. The GTCC countermeasure \cite{valero2012gammatone} obtained the highest EER of 55.0. The results demonstrate the inadequacy of the countermeasures in testing the speech sample with the artifacts of distinct microphonic discrepancies. Except for the LFCC \cite{zhou2011linear} countermeasure, none of the countermeasures achieved an EER lower than 40.0. The detailed countermeasure results of this experiment are presented in Table \ref{tab:GMMvsdc}.

From this experiment, we can infer that the countermeasures are totally dependent on the dataset characteristics. In the ASVspoof2019 dataset, for instance, the APGDF \cite{rajan2013using} features produced the best results. On the VSDC dataset, however, the APGDF \cite{rajan2013using} features failed to generate effective outcomes. Similarly, LPCC  \cite{wong2001comparison} did not generate effective results for the ASVspoof datasets, particularly in the ASVspoof-PA assessment set, but produced the best results in VSDC testing. Furthermore, we can see that the front end features that worked well against logical attacks did not perform well against physical attacks. The VSDC dataset also includes PA (replay) attacks that resulted in the performance degradation of each countermeasure. However, despite the dataset's wide variability, the APGDF \cite{rajan2013using}, CQCC \cite{Todisco2017ConstantQC}, and LPCC  \cite{wong2001comparison} features produce comparable results independent of the dataset's specific configuration.

\subsection{Performance analysis of SOTA countermeasures with an SVM classifier}
In this experiment, we evaluate the effectiveness of SOTA countermeasures against the SVM classifier. The SVM classifier has been shown to be a superior optimal classifier to the GMM classifier \cite{9107388}. An SVM with an RBF distribution was utilized in the experiment. The RBF distribution was selected after a thorough analysis of all feasible distributions. The scikit-learn \cite{scikit-learn} machine learning library was used to train and test the SVM classifier. In addition, the classification report \cite{scikit-learn} was used to compute the precision, recall, F1-score, and accuracy of the countermeasures. The results of this experiment on the VSDC dataset are mentioned in Table \ref{tab:svmvsdc}, and demonstrate that the CQCC \cite{Todisco2017ConstantQC} countermeasure proved to have the most robust front-end attributes in comparison with the SOTA countermeasures. CQCC \cite{Todisco2017ConstantQC} features, in particular, achieved the lowest EER of 0.27, the highest precision, recall, and F1-scores of 0.92, 0.85, and 0.88, respectively, as well as the highest accuracy of 0.93. Spectral features such as SCFC \cite{kua2010investigationOS}, on the other hand, performed poorly in the case of the VSDC dataset, which contains audio samples with micro-phonic inconsistencies. For instance, the SCFC \cite{kua2010investigationOS} features had the lowest precision (0.39), recall (0.50), and F1-score (0.44), and achieved an accuracy of 0.78. Similarly, the SCMC \cite{kua2010investigationOS} features were found to be the second worst classifier, after the SCFC \cite{kua2010investigationOS}, with a 0.53 EER. This indicates that the spectral centroid magnitude and frequency-based coefficients failed to perform effectively when micro-phonic distinction (i.e., VSDC) exists in the speech sample. In contrast, the constant Q-cepstral-based coefficient successfully overcame the microphone's fluctuation.

In the case of the ASVspoof2019 dataset, each of the SOTA countermeasures was tested against the SVM classifier, along with the development and evaluation set of the datasets, and the results are reported in Table \ref{tab:SVMASV}. The results demonstrate that in the testing of LA attacks that exist in the development set and evaluation set, the CQCC \cite{Todisco2017ConstantQC} and IMFCC \cite{chakroborty2009improved} countermeasures achieved comparable performance, with an EER of 0.30 and 0.29, respectively. Similarly, for the evaluation set, the CQCC \cite{Todisco2017ConstantQC} and IMFCC \cite{chakroborty2009improved} countermeasures achieved the best EER of 0.69 and 0.76, respectively. Although the EER of 0.69 and 0.76 was not good enough for any spoofing system, none of the other countermeasures achieved an EER lower than these countermeasures.
From this experiment, it was concluded that the performance of the majority of the spectral and cepstral countermeasures declined when tested against different classifiers with different configurations. However, the CQCC \cite{Todisco2017ConstantQC} countermeasure performed far better than the other countermeasures. In addition, in the presence of micro-phonic distinctions, CQCC \cite{Todisco2017ConstantQC} features work effectively, while the rest of the spectral features fail to perform adequately. Furthermore, the phase-based APGDF \cite{rajan2013using} features were the third best features compared to the state-of-the-art that achieved the optimal results. Detailed results of the SOTA countermeasures are presented in Table \ref{tab:SVMASV}.

Based on the experiments involving the SOTA countermeasures with machine learning-based GMM and SVM-based classifiers, it is determined that the CQCC \cite{Todisco2017ConstantQC} is the best countermeasure against voice spoofing attacks. More specifically, in the testing of replay attacks, where the performance of the state-of-the-art classifiers drops significantly, the CQCC \cite{Todisco2017ConstantQC} features scored the optimal EER of 0.27, 0.43, and 0.30 in GMM and SVM testing, respectively. The CQCC \cite{Todisco2017ConstantQC} countermeasure achieved the lowest EER and min-TDCF when in development. The results of both ML-based classifiers show that the phase-based APGDF \cite{rajan2013using}, linear LFCC \cite{zhou2011linear} , and inverted Mel-based IMFCC \cite{chakroborty2009improved} countermeasures performed better than the comparative methods in all experiments and received the optimal scores in all performance measurements. APGDF \cite{rajan2013using} also had the lowest EER and min-TDCF in the ASVspoof2019-PA dataset evaluation. Similar results are obtained with the VSDC and SVM classifiers. In comparison to the comparative approaches, the SCFC \cite{kua2010investigationOS} features were shown to be the weakest. SCFC \cite{kua2010investigationOS} failed to perform well in any experiment, and had the highest EER and min-TDCF. Although the CQCC \cite{Todisco2017ConstantQC}  and APGDF  \cite{rajan2013using} front end features clearly outperformed all comparable features in the ASVspoof2019-LA, ASVspoof2019-PA, and VSDC datasets, the question of performance when the ASV encounters unfamiliar samples from unknown surroundings and devices persists. Although the ML-based GMM and SVM classifiers showed the capabilities of the countermeasure, they have become the conventional classifiers for spoof detection. In contrast, several deep learning and end-to-end solutions have been published recently to differentiate between spoofed and bonafide speech samples. Therefore, we tested the countermeasures against the recent CNN and CNN-GRU based classifiers and also performed cross-corpus experiments in order to validate the generalized performance of both the highest performing countermeasures. The results are discussed in the next subsection.

\begin{table}[!t]
\caption{Experimental performance of the countermeasures with VSDC dataset and SVM based classifier.}
\label{tab:svmvsdc}
\centering
\begin{tabular}{*{6}{c}}
\hline
Feature & EER   & Precision &  Recall & F1-score & Accuracy\\ 
 \hline
LFCC \cite{zhou2011linear}   &  0.49  &  0.88     & 0.74      &  0.79      & 0.88  \\
 \hline
CQCC \cite{Todisco2017ConstantQC}   &  0.27  &  0.92     & 0.85      &  0.88      & 0.93  \\
 \hline
LPCC  \cite{wong2001comparison}  &  0.44  &  0.88     & 0.77      &  0.81      & 0.89 \\
 \hline
PSRCC  \cite{tapkir2018novel} &  0.40  &  0.79     & 0.76      &  0.77      &  0.86 \\
 \hline
MSRCC  \cite{tapkir2018novel}  &  0.46  &  0.84     & 0.76      &  0.79      &  0.86 \\
 \hline
SCFC  \cite{kua2010investigationOS}  &  0.80  &  0.39     & 0.50      &  0.44      & 0.78 \\
 \hline
SCMC \cite{kua2010investigationOS}   &  0.53  &  0.87     & 0.72      &  0.76      & 0.87  \\
 \hline
MFCC \cite{muda2010voice}    &  0.46 &   0.82     & 0.75      &  0.77      & 0.86  \\
 \hline
IMFCC \cite{chakroborty2009improved}  &  0.71  &  0.88     & 0.64      &  0.67      &  0.84 \\ 
\hline
RPS  \cite{saratxagainproceedings}   &  0.40  &  0.76     & 0.78      &  0.77      &  0.85 \\
 \hline
RFCC \cite{saratxagainproceedings}   &  0.51  &  0.87     & 0.73      &  0.76      &  0.87 \\
 \hline
GTCC  \cite{valero2012gammatone}   &  0.42  &  0.82     & 0.76      &  0.78      &  0.86 \\
 \hline
APGDF  \cite{rajan2013using}  &  0.43  &  0.89     & 0.77      &  0.81      &  0.89  \\
 \hline
SSFC \cite{tapkir2018novel}  &  0.44   &  0.87     & 0.77      &  0.80      &  0.88  \\
 \hline
\end{tabular}
\end{table}

\begin{table}[!t]
\caption{Experimental performance of the countermeasures with ASVspoof2019 and an SVM based classifier.}
\label{tab:SVMASV}
\centering
\begin{tabular}{*{6}{c}}
\hline
 \multirow{2}{1.5cm}{Feature} & \multirow{2}{1cm}{Dataset}  & \multicolumn{2}{c}{Development} & \multicolumn{2}{c}{Evaluation} \\ 
                        &   & EER & Acc & EER & Acc  \\
 \hline
 \multirow{2}{1.5cm}{LFCC \cite{zhou2011linear}}  & ASVspoof19-LA &  0.58 & 0.90 & 0.91 & 0.12  \\
                        & ASVspoof19-PA & 0.72 & 0.75 & 0.83 & 0.11 \\
 \hline
 \multirow{2}{1.5cm}{CQCC \cite{Todisco2017ConstantQC}} & ASVspoof19-LA & 0.30 & 0.94 & 0.69 & 0.29  \\
                          & ASVspoof19-PA & 0.42 & 0.29 & 0.80  & 0.19  \\
 \hline
 \multirow{2}{1.5cm}{LPCC \cite{wong2001comparison}} & ASVspoof19-LA & 0.50 & 0.88 & 0.80 & 0.15 \\
                        & ASVspoof19-PA & 0.61 & 0.20 & 0.81 & 0.19  \\
 \hline
 \multirow{2}{1.5cm}{SCFC \cite{kua2010investigationOS}}& ASVspoof19-LA & 0.50 & 0.89 &  0.90 & 0.16 \\
                        & ASVspoof19-PA & 0.75 & 0.19 & 0.89 & 0.14\\
 \hline
 \multirow{2}{1.5cm}{SCMC \cite{kua2010investigationOS}}& ASVspoof19-LA &0.49 &0.93& 0.88  & 0.12  \\
                        & ASVspoof19-PA & 0.75 & 0.20 &  0.86 &  0.15 \\
 \hline
 \multirow{2}{1.5cm}{PSRCC \cite{tapkir2018novel}}& ASVspoof19-LA & 0.49 & 0.89 &  0.89 &  0.11 \\
                        & ASVspoof19-PA & 0.84 & 0.25&  0.89 &  0.12\\
 \hline
 \multirow{2}{1.6cm}{MSRCC \cite{tapkir2018novel}}& ASVspoof19-LA & 0.51 & 0.90 &  0.91 &  0.11 \\
                        & ASVspoof19-PA & 0.84 & 0.25&  0.89 &  0.12\\
 \hline
 \multirow{2}{1.5cm}{MFCC \cite{muda2010voice}}& ASVspoof19-LA & 0.50 & 0.89 & 0.80 & 0.17 \\
                        & ASVspoof19-PA & 0.76 & 0.22 & 0.86 & 0.16 \\
 \hline
 \multirow{2}{1.5cm}{IMFCC \cite{chakroborty2009improved}}& ASVspoof19-LA & 0.29 &0.94  & 0.76  &  0.24 \\
                        & ASVspoof19-PA & 0.51 & 0.26 &  0.82 &  0.18 \\
 \hline
 \multirow{2}{1.5cm}{RFCC \cite{saratxagainproceedings}}& ASVspoof19-LA & 0.46 &0.93 &  0.87 &  0.14 \\
                        & ASVspoof19-PA & 0.84 & 0.26 & 0.88  &  0.14\\
 \hline
 \multirow{2}{1.5cm}{RPS \cite{saratxagainproceedings}} & ASVspoof19-LA & 0.52 & 0.90 &  0.91 &  0.10 \\
                        & ASVspoof19-PA & 0.86 & 0.25 & 0.89  &  0.11 \\
 \hline
 \multirow{2}{1.5cm}{SSFC \cite{tapkir2018novel}}& ASVspoof19-LA & 0.50 & 0.90 &  0.89 &  0.12 \\
                        & ASVspoof19-PA & 0.87 & 0.29 & 0.90  &  0.10 \\
 \hline
 \multirow{2}{1.5cm}{GTCC \cite{valero2012gammatone}}& ASVspoof19-LA & 0.50 & 0.90 &  0.90 &  0.10 \\
                        & ASVspoof19-PA & 0.84 & 0.25&  0.89 &  0.12\\
 \hline
\multirow{2}{1.5cm}{APGDF \cite{rajan2013using}}& ASVspoof19-LA &  0.34 & 0.94 & 0.82  & 0.17  \\
                        & ASVspoof19-PA & 0.60 & 0.17& 0.83  &  0.15  \\
 \hline
\end{tabular}
\end{table}

\subsection{Performance analysis of the SOTA countermeasures with a CNN-based classifier}
In this subsection, we show the effectiveness of SOTA countermeasures along with the DL-based CNN and CNN-GRU classifier. The CNN classifier is one of the most recent and most advanced neural network-based classifiers for the detection of voice spoofing on ASV systems. The SOTA countermeasures were tested against the CNN classifier, and the results are shown in Table \ref{tab:cnnaloneresults}.  The same architectural CNN was used in the experiments that was adopted in \cite{jung2019replay}. The open source code is implemented in the mentioned configuration in section VIII subsection C.

\begin{table}[!b]
\caption{Experimental performance analysis of the countermeasures on the training and testing subsets of the same dataset with a CNN based classifier.}
\label{tab:cnnaloneresults}
\centering
\begin{tabular}{*{5}{c}}
 \hline
Features & Training& Testing &  EER &min-tdcf\\
   \hline
 \multirow{2}{2cm}{IMFCC \cite{chakroborty2009improved}} & VSDC   & VSDC     & 0.06 &  0.17 \\
  & ASVspoof19   & ASVspoof19   & 0.19 &  0.50 \\ 
  \hline
 \multirow{2}{2cm}{SSFC\cite{tapkir2018novel}} & VSDC   & VSDC     & 0.03 &  0.07 \\
 & ASVspoof19   & ASVspoof19   &  0.18 & 0.46  \\ 
  \hline
 \multirow{2}{2cm}{RFCC \cite{saratxagainproceedings}} & VSDC   & VSDC     &  0.13 & 0.25  \\
 & ASVspoof19   & ASVspoof19   & 0.05 &  0.13 \\ 
   \hline
 \multirow{2}{2cm}{SCFC \cite{kua2010investigationOS}} & VSDC   & VSDC     & 0.10 & 0.30  \\
& ASVspoof19   & ASVspoof19 & 0.14 & 0.36  \\
  \hline
 \multirow{2}{2cm}{SCMC\cite{kua2010investigationOS}}& VSDC   & VSDC     & 0.14 & 0.37  \\
 & ASVspoof19   & ASVspoof19 & 0.44 &  0.94\\
    \hline
 \multirow{2}{2cm}{PSRCC \cite{tapkir2018novel}} & VSDC   & VSDC     & 0.08 & 0.10  \\
 & ASVspoof19   & ASVspoof19   & 0.43 & 0.75  \\ 
 \hline 
 \multirow{2}{2cm}{MSRCC \cite{tapkir2018novel}} & VSDC   & VSDC     & 0.13 & 0.17  \\
 & ASVspoof19   & ASVspoof19   & 0.43 & 0.75  \\ 
 \hline
 \multirow{2}{2cm}{RPS \cite{saratxagainproceedings}} & VSDC   & VSDC     & 0.04 & 0.11  \\
& ASVspoof19   & ASVspoof19     & 0.21 &  0.65 \\
\hline
\multirow{2}{2cm}{CQCC \cite{Todisco2017ConstantQC}}  & VSDC   & VSDC     & 0.04 &   0.13\\
 & ASVspoof19   & ASVspoof19   & 0.17 & 0.45  \\ 
 \hline
\multirow{2}{2cm}{LFCC \cite{zhou2011linear}}  & VSDC   & VSDC     & 0.03 & 0.03 \\
 & ASVspoof19   & ASVspoof19   & 0.16 &  0.43 \\ 
  \hline
 \multirow{2}{2cm}{MFCC \cite{muda2010voice}}  & VSDC   & VSDC     & 0.06 & 0.16  \\
 & ASVspoof19   & ASVspoof19   & 0.26  & 0.62 \\ 
  \hline
 \multirow{2}{2cm}{LPCC \cite{wong2001comparison}} & VSDC   & VSDC     & 0.02 &  0.05 \\
 & ASVspoof19   & ASVspoof19   & 0.19 &  0.50 \\ 
  \hline
 \multirow{2}{2cm}{GTCC \cite{valero2012gammatone}} & VSDC   & VSDC     & 0.03 & 0.07  \\
 & ASVspoof19   & ASVspoof19   & 0.43 & 0.75  \\ 
 \hline
\multirow{2}{2cm}{APGDF \cite{rajan2013using}}& VSDC   & VSDC     & 0.01 & 0.04 \\
 & ASVspoof19   & ASVspoof19   & 0.17 & 0.45  \\ 
\hline
 \end{tabular}
\end{table}

The results for the VSDC dataset demonstrate that the APGDF \cite{rajan2013using} front end features outperform the SOTA countermeasures with an EER of 0.01 and a min T-DCF of 0.04. Following all of the countermeasures, the linear-based features perform well with the CNN-based classifier to detect voice spoofing attacks. For instance, according to performance analysis with EER, LPCC \cite{wong2001comparison} and LFCC \cite{zhou2011linear} rank second and third, with EERs of 0.026 and 0.032, respectively. Similarly, in the case of ASVspoof19, the rectangular-based RFCC \cite{saratxagainproceedings} features outperform all evaluated countermeasures on the ASVspoof19 dataset, with an EER of 0.054 and a min T-DCF of 0.139. The magnitude-based feature was found to be significantly deficient during the ASVspoof19 evaluation. For instance, MSRCC \cite{tapkir2018novel} features had the highest EER and min-TDCF values, at 0.443 and 0.947, respectively. In contrast, the highest EER of 0.21, and the highest min-TDCF score of 0.43, were seen in the SCFC \cite{kua2010investigationOS} features against VSDC. Moreover, when compared to the GMM classifier, the CNN classifier performed significantly better across the board. For instance, the highest EER recorded with the GMM classifier was 0.80 on the LFCC feature, while the highest EER achieved with the CNN classifier was just 0.265 with the MFCC \cite{muda2010voice} features. Using the same feature set but replacing the classifier, the EER was reduced by nearly 65\%. This experiment clearly shows the significance of back end classifiers in speech spoofing detection. Detailed results of the SOTA countermeasures are presented in Table \ref{tab:cnnaloneresults}.

\subsection{Experimental Analysis of the SOTA countermeasures with a CNN-GRU based classifier}
In this experiment, a CNN-GRU \cite{jung2019replay} classifier was used to classify the SOTA countermeasures in order to test their effectiveness against the SS, VC, and replay attacks, and the results of this experiment are presented in Table XVIII. The results demonstrate that in the case of the VSDC dataset, the linear and spectral magnitude-based LFCC and SCMC features achieved a comparatively similar performance, with an EER of 0.02 and a min T-DCF of 0.05. Due to the versatility of the VSDC dataset, the SCFC features scored an EER of 0.18 and a min T-DCF of 0.35. In particular, the results from this experiment proved to be optimal for the VSDC dataset, in comparison with the GMM, SVM, and CNN-GRU classifiers. Moreover, the SOTA countermeasures achieved the best result on the CNN-GRU classifiers. In contrast, in the case of ASVspoof19, LPCC was the best performing countermeasure, with an EER of 0.02 and a min T-DCF of 0.47, while the MFCC and MSRCC countermeasures have the lowest EERs of 0.22 and 0.18, respectively. Therefore, it was concluded from this experiment that the performance of the existing SOTA countermeasure differs in the presence of distinct classifiers and may not function adequately in the presence of advanced spoofing attempts. In addition, we can infer that the CNN-GRU classifier proved to be the optimal classifier for both of the datasets. However, the effectiveness of the SOTA countermeasure across corpora must still be demonstrated. Therefore, in the next section we provide a cross-corpus evaluation of the SOTA countermeasures. 

\begin{table}[!t]
\caption{Experimental performance analysis of the SOTA countermeasures with a CNN-GRU based classifier.}
\label{tab:cnngrualoneresults}
\centering
\begin{tabular}{*{5}{c}}
\hline
Feature & Train Dataset & Test Dataset &  EER &min T-DCF\\
   \hline
\multirow{2}{1.5cm}{LFCC \cite{zhou2011linear}} & VSDC   & VSDC     & 0.02 &  0.05 \\
 & ASVspoof19   & ASVspoof19   & 0.15 & 0.41 \\ 
  \hline
 \multirow{2}{1.5cm}{SSFC\cite{tapkir2018novel}} & VSDC   & VSDC     & 0.04 & 0.07  \\
 & ASVspoof19   & ASVspoof19   & 0.17  & 0.46 \\
 \hline
 \multirow{2}{1.5cm}{MFCC \cite{muda2010voice}} & VSDC   & VSDC  & 0.03 & 0.07  \\
  & ASVspoof19   & ASVspoof19   & 0.22 & 0.548 \\
  \hline
\multirow{2}{1.5cm}{SCFC\cite{kua2010investigationOS}} & VSDC   & VSDC     & 0.18 & 0.35  \\
  & ASVspoof19   & ASVspoof19   & 0.09 & 0.25 \\
 \hline
 \multirow{2}{1.5cm}{LPCC \cite{wong2001comparison}} & VSDC   & VSDC     & 0.18& 0.03  \\
  & ASVspoof19   & ASVspoof19   & 0.02 & 0.47 \\ 
 \hline
 \multirow{2}{1.5cm}{SCMC\cite{kua2010investigationOS}} & VSDC   & VSDC     & 0.02 & 0.05  \\
 & ASVspoof19   & ASVspoof19   & 0.17  & 0.46 \\
  \hline
 \multirow{2}{1.6cm}{MSRCC \cite{tapkir2018novel}} & VSDC   & VSDC     & 0.21 &  0.03 \\
& ASVspoof19   & ASVspoof19     & 0.22 & 0.50  \\
 \hline
 \multirow{2}{1.5cm}{PSRCC \cite{tapkir2018novel}} & VSDC   & VSDC     & 0.23 &  0.03 \\
& ASVspoof19   & ASVspoof19     & 0.19 & 0.53  \\
  \hline
\multirow{2}{1.5cm}{RFCC \cite{saratxagainproceedings}} & VSDC   & VSDC     & 0.11 &  0.22 \\
  & ASVspoof19   & ASVspoof19   & 0.04 & 0.11 \\ 
 \hline
 \multirow{2}{1.5cm}{IMFCC\cite{chakroborty2009improved}} & VSDC   & VSDC     & 0.03 & 0.09  \\
 & ASVspoof19   & ASVspoof19   & 0.18 & 0.50 \\ 
 \hline
\multirow{2}{1.5cm}{APGDF \cite{rajan2013using}} & VSDC   & VSDC     & 0.08 & 0.26  \\
 & ASVspoof19   & ASVspoof19     & 0.14 & 0.37\\
 \hline
\multirow{2}{1.5cm}{CQCC \cite{Todisco2017ConstantQC}}  & VSDC   & VSDC     & 0.09 &  0.23 \\
 & ASVspoof19   & ASVspoof19     & 0.18& 0.49  \\
 \hline
 \multirow{2}{1.5cm}{GTCC \cite{valero2012gammatone}} & VSDC   & VSDC     & 0.04 &  0.09 \\
& ASVspoof19   & ASVspoof19     & 0.19 & 0.53  \\
  \hline
  \multirow{2}{1.5cm}{RPS \cite{saratxagainproceedings}}& VSDC   & VSDC     & 0.04 &  0.09 \\
& ASVspoof19   & ASVspoof19     & 0.19 & 0.53  \\
 \hline
\end{tabular} 
\end{table}

\subsection{Cross-corpus evaluations of SOTA countermeasures}
The SOTA countermeasures are compared in this section using cross-corpus evaluation. Cross-corpus evaluation and generalisation was one of the key needs mentioned by the in-field researcher \cite{tan2021survey}, which we resolved in this experiment. A key shortcoming in previous research is the lack of cross-corpus experiments \cite{sahidullah2019introduction}. Cross-corpus examination demonstrates the countermeasure's effectiveness in dealing with unknown aspects of spoofing attacks by removing prior bias from the model in the form of characteristics shared by a particular dataset, and allows it to perform a classification based solely on how the model recognises the unique aspects of spoofing attacks. In addition, this experiment was performed to determine the comprehensiveness of the countermeasures to speech samples, including unknown and diverse spoofing artifacts such as distinct VC algorithms, SSD techniques, and others. During the experiment, the countermeasures were trained using one dataset and the testing was done using the speech samples from another. There are some similarities between the two datasets. Replayed speech samples, for instance, can be found in the VSDC and ASVspoof19 databases. Initially, the countermeasures were trained using VSDC data samples and then tested with the replayed spoof speech samples from the ASVspoof corpus. The SOTA countermeasure was then trained using ASVspoof and evaluated using VSDC data samples. This evaluation was performed against three distinct classifiers, and the results are shown below in Tables \ref{tab:gmmcrossresults}, \ref{tab:cnncrossresults}, and \ref{tab:cnngrualoneresults}. 

\subsubsection{Cross-corpus evaluation of the SOTA countermeasures using a GMM classifier}
In this subsection, performance of the SOTA countermeasure was against a GMM classifier, and the results are reported in Table \ref{tab:gmmcrossresults}. The results demonstrate that when the classifier was trained on the VSDC dataset and tested on the ASVspoof19 dataset, the countermeasure PSRCC \cite{tapkir2018novel} outperformed the SOTA countermeasures, with an EER of 0.25 and a min TDCF of 0.39. When trained on the ASVspoof19 dataset and evaluated on the VSDC dataset, RFCC \cite{saratxagainproceedings} was the top performing countermeasure, with the lowest EER of 0.39 and the lowest min T-DCF of 0.27. In contrast, when training on VSDC and testing with the ASVspoof dataset, the countermeasure SSFC \cite{kua2010investigationOS} obtained the highest EER of 0.80 and min-tdcf of 0.99. Similarly, the SSFC \cite{tapkir2018novel} countermeasure obtained the highest EER of 0.67. It is observed from the results that the EER of the SOTA countermeasures varies from 0.25 to 0.80, but min-TDCF ranges from 0.81 to 1.0. Only the PSRCC \cite{tapkir2018novel}, MSRCC\cite{tapkir2018novel}, RFCC \cite{saratxagainproceedings}, GTCC \cite{valero2012gammatone}, and APGDF \cite{rajan2013using} countermeasures are able to achieve lower min-TDCF, at 0.39, 0.47, 0.27, 0.38, and 0.59, respectively. The EER performance metric is used to evaluate the effectiveness of spoofing countermeasures, whereas the min-TDCF evaluation parameter can be used to evaluate the efficacy of ASVs. Consequently, based on these results, we may conclude that the effectiveness of the SOTA countermeasures will be called into question when employed for ASV evaluations. 

Surprisingly, no specific SOTA countermeasure performed considerably better in both cross-corpus evaluations. For instance, when trained using ASVspoof19 and evaluated against the VSDC dataset, the countermeasure MSRCC worked effectively. In the ASVspoof dataset test, EER increased by up to 11\%. Despite being the worst performer in the standalone tests, PSRCC performed well when trained on VSDC and evaluated with ASVspoof19. While phase-based spectral features outperformed the state-of-the-art in cross-corpus evaluation, the current investigation revealed that when utilizing the GMM classifier, the performance of the existing countermeasures differed greatly in cross-corpus evaluation. Moreover, the results show that when the system was unfamiliar with the testing speech samples, the countermeasures failed to accurately recognize the spoofed speech samples. However, determining the countermeasure's cross-corpus performance using a single classifier is not practical for fairness and may be biased. Therefore, along with the GMM classifier, we demonstrate the efficacy of the SOTA countermeasures against the DL-based CNN and CNN-GRU classifiers, and the results are reported below.

\begin{table}[!b]
\caption{Cross corpus evaluation of the SOTA countermeasures with GMM.}
\label{tab:gmmcrossresults}
\centering
\begin{tabular}{*{5}{c}}
 \hline
Features & Training& Testing &  EER &min-tdcf\\
 \hline
\multirow{2}{2cm}{CQCC \cite{Todisco2017ConstantQC}} & VSDC   & ASVspoof19  & 0.40 &   0.81\\
  & ASVspoof19   & VSDC    & 0.66 & 0.99\\
 \hline
\multirow{2}{2cm}{LFCC \cite{zhou2011linear}}  & VSDC & ASV'19 PA & 0.60 & 0.99 \\
 & ASV'19 PA & VSDC & 0.45& 0.99\\
 \hline
  \multirow{2}{2cm}{SSFC \cite{tapkir2018novel}} & VSDC   & ASVspoof19     & 0.80 & 0.99  \\
  & ASVspoof19   & VSDC     & 0.67 & 0.99  \\
 \hline
 \multirow{2}{2cm}{RPS \cite{saratxagainproceedings}} & VSDC   & ASVspoof19     & 0.50 & 0.93  \\
  & ASVspoof19   & VSDC     & 0.40 & 0.94  \\
 \hline
 \multirow{2}{2cm}{LPCC \cite{wong2001comparison}} & VSDC   & ASVspoof19     & 0.41 & 0.89  \\
  & ASVspoof19   & VSDC     & 0.62 & 1.0  \\
\hline
 \multirow{2}{2cm}{PSRCC \cite{tapkir2018novel}} & VSDC   & ASVspoof19     & 0.25 & 0.39  \\
 & ASVspoof19   & VSDC   & 0.67 & 0.99  \\
 \hline 
 \multirow{2}{2cm}{MSRCC \cite{tapkir2018novel}} & VSDC   & ASVspoof19     & 0.29 & 0.47  \\
 & ASVspoof19   & VSDC   & 0.40 & 0.94   \\
 \hline
 \multirow{2}{2cm}{SCFC \cite{kua2010investigationOS}} & VSDC   & ASVspoof19   & 0.58& 0.99  \\
 & ASVspoof19   & VSDC    & 0.48 & 0.99  \\
 \hline
 \multirow{2}{2cm}{SCMC\cite{kua2010investigationOS}} & VSDC & ASVspoof19   & 0.60  & 0.97  \\
 & ASVspoof19   & VSDC    & 0.46 & 0.87  \\
 \hline
 \multirow{2}{2cm}{MFCC \cite{muda2010voice}} & VSDC   & ASVspoof19   & 0.42 & 0.97  \\
 & ASVspoof19   & VSDC    & 0.45 & 0.93  \\
 \hline
 \multirow{2}{2cm}{IMFCC \cite{chakroborty2009improved}} & VSDC   & ASVspoof19     & 0.45 & 0.99  \\
 & ASVspoof19   & VSDC     & 0.45 & 0.96  \\
 \hline
 \multirow{2}{2cm}{RFCC \cite{saratxagainproceedings}}& VSDC   & ASVspoof19     & 0.34 & 0.84  \\
 & ASVspoof19   & VSDC   & 0.39 & 0.27   \\
 \hline
 \multirow{2}{2cm}{GTCC \cite{valero2012gammatone}} & VSDC   & ASVspoof19     & 0.36 & 0.38  \\
 & ASVspoof19   & VSDC   & 0.56 & 1.0   \\
 \hline
\multirow{2}{2cm}{APGDF \cite{rajan2013using}} & VSDC   & ASVspoof19     & 0.30 &  0.59  \\
 & ASVspoof19   & VSDC   & 0.61  &  0.99  \\
 \hline
 \end{tabular}
\end{table}

\subsubsection{Cross corpus evaluation of SOTA countermeasures with a CNN based classifier}
In this section, cross-corpus evaluations of the SOTA countermeasures with a CNN-based classifier are discussed, and the results are presented in Table \ref{tab:cnncrossresults}. The results demonstrate that in VSDC training and ASVspoof19 testing, the phase-oriented PSRCC \cite{tapkir2018novel} countermeasure performed the best and obtained an EER of 0.37 and a min T-DCF of 0.94. In contrast, in the case of ASVspoof19 training and VSDC testing, the SCMC \cite{kua2010investigationOS} features achieved the minimum values for EER and min T-DCF, at 0.35 and 0.96, respectively. Although the countermeasures and the CNN-based classifier performed marginally better when trained with VSDC and tested with the ASVspoof19 dataset, several of the front-end features performed significantly worse. This uncertainty shows that the performance of these countermeasures remains ambiguous against unseen attacks without having prior knowledge of the speech sample.

\begin{table}[!t]
\caption{Cross corpus evaluation of the SOTA countermeasures with a CNN based classifier.}
 \label{tab:cnncrossresults}
\centering
\begin{tabular}{*{5}{c}}
\hline
Feature & Train Dataset & Test Dataset &  EER &min T-DCF\\
  \hline
 \multirow{2}{1.5cm}{IMFCC \cite{chakroborty2009improved}} & VSDC  & ASVspoof19   & 0.43 & 0.99  \\
  & ASVspoof19   & VSDC    & 0.57 &  1.0 \\ 
  \hline
 \multirow{2}{1.5cm}{SCMC \cite{kua2010investigationOS}} & VSDC   & ASVspoof19   &  0.51  & 0.99  \\
 & ASVspoof19   & VSDC    & 0.35  &  0.96 \\
  \hline
\multirow{2}{1.5cm}{RFCC \cite{saratxagainproceedings}} & VSDC   & ASVspoof19   & 0.40  &  0.96\\
 & ASVspoof19   & VSDC    & 0.50 & 0.91  \\
   \hline
 \multirow{2}{1.5cm}{RPS \cite{saratxagainproceedings}} & VSDC   & ASVspoof19    & 0.68 & 0.99  \\
& ASVspoof19   & VSDC     & 0.55 & 0.98  \\
  \hline
\multirow{2}{1.5cm}{SSFC \cite{tapkir2018novel}} & VSDC   & ASVspoof19     & 0.72 &  0.99 \\
 & ASVspoof19   & VSDC     & 0.39 & 0.91\\
  \hline
 \multirow{2}{1.5cm}{PSRCC \cite{tapkir2018novel}} & VSDC   & ASVspoof19   & 0.37 &  0.94 \\
 & ASVspoof19   & VSDC    & 0.51 &  0.97\\
  \hline
 \multirow{2}{1.6cm}{MSRCC \cite{tapkir2018novel}} & VSDC   & ASVspoof19   & 0.46 & 0.52 \\
 & ASVspoof19   & VSDC    & 0.47 & 0.93\\
 
  \hline
\multirow{2}{1.5cm}{CQCC \cite{Todisco2017ConstantQC}}  & VSDC   & ASVspoof19   & 0.48 &   0.98\\
& ASVspoof19   & VSDC    & 0.66 & 1.0  \\
 \hline
\multirow{2}{1.5cm}{LFCC \cite{zhou2011linear}}  & VSDC   & ASVspoof19   & 0.51 &  0.99 \\
 & ASVspoof19   & VSDC    & 0.37 & 0.99  \\
  \hline
 \multirow{2}{1.5cm}{MFCC \cite{muda2010voice}}  & VSDC   & ASVspoof19   & 0.47 & 0.93  \\
& ASVspoof19   & VSDC    & 0.52 &  0.94\\
   \hline
 \multirow{2}{1.5cm}{SCFC\cite{kua2010investigationOS}} & VSDC   & ASVspoof19   & 0.53 & 0.99  \\
 & ASVspoof19   & VSDC    & 0.54 &  0.97  \\
  \hline
 \multirow{2}{1.5cm}{LPCC \cite{wong2001comparison}} & VSDC   & ASVspoof19   & 0.46 & 0.98  \\
 & ASVspoof19   & VSDC    & 0.63 &  1.0 \\
  \hline
 \multirow{2}{1.5cm}{GTCC \cite{valero2012gammatone}} & VSDC   & ASVspoof19   & 0.50 & 0.98  \\
 & ASVspoof19   & VSDC    & 0.41 & 0.91  \\
 \hline
\multirow{2}{1.5cm}{APGDF \cite{rajan2013using}} & VSDC   & ASVspoof19   & 0.44 &  0.99 \\
& ASVspoof19   & VSDC    & 0.56 &  1.0 \\
\hline
 \end{tabular}
\end{table}

\subsubsection{Cross corpus measurements of the SOTA countermeasures with a CNN-GRU based classifier}
A CNN-GRU classifier was used to evaluate the countermeasures across corpora in this subsection, and the results are shown in Table \ref{tab:cnncrossresults}. In the instance of VSDC training and ASVspoof19 testing, the APGDF-based countermeasure \cite{rajan2013using} obtained an optimal EER of 0.34 and a minimum T-DCF score of 0.82. With an EER of 0.44, the RFCC \cite{saratxagainproceedings} and IMFCC \cite{chakroborty2009improved} were ranked second and third, respectively. In the case of ASVspoof19 training and VSDC testing, the MFCC-based countermeasure \cite{muda2010voice} outperformed the SOTA countermeasures with the lowest EER of 0.34 and the lowest min T-DCF score of 0.97. The outcomes of this experiment were similar to the results of the CNN-based cross-corpus evaluations. The min-TDCF of the SOTA countermeasures was greater than the countermeasures' standalone testing. In the instance of ASVspoof19 dataset training and VSDC testing, the results demonstrated that the majority of the countermeasures' minimum T-DCF was at or around 1. These findings demonstrate the ineffectiveness of existing countermeasures and classifiers against unknown spoofing attacks. Furthermore, these findings have raised concerns regarding the effectiveness of existing protections against complicated contemporary spoofing attacks.

\begin{table}[!t]
\caption{Cross corpus evaluation of the countermeasures with a CNN-GRU based classifier.}
\label{tab:cnngrualoneresults}
\centering
\begin{tabular}{*{5}{c}}
 \hline
 Feature & Train Dataset & Test Dataset &  EER &min T-DCF\\
\hline
\multirow{2}{1.5cm}{LFCC \cite{zhou2011linear}}  & VSDC   & ASVspoof19   & 0.47 &  0.99 \\
  & ASVspoof19   & VSDC    & 0.53 &  0.98 \\
 \hline
 \multirow{2}{1.5cm}{MFCC \cite{muda2010voice}}  & VSDC   & ASVspoof19   & 0.45  & 0.94  \\
 & ASVspoof19   & VSDC    & 0.34 &  0.97 \\
  \hline
\multirow{2}{1.5cm}{SCFC\cite{kua2010investigationOS}}  & VSDC   & ASVspoof19   & 0.49 & 0.99  \\
 & ASVspoof19   & VSDC    & 0.59 & 0.99  \\
 \hline
 \multirow{2}{1.5cm}{LPCC \cite{wong2001comparison}}  & VSDC   & ASVspoof19   & 0.65 & 0.99  \\
 & ASVspoof19   & VSDC    & 0.45 & 1.0  \\
 \hline
 \multirow{2}{1.5cm}{SCMC\cite{kua2010investigationOS}}  & VSDC   & ASVspoof19   & 0.45 & 0.99  \\
 & ASVspoof19   & VSDC    & 0.65 & 1.0 \\
  \hline
 \multirow{2}{1.5cm}{PSRCC \cite{tapkir2018novel}} & VSDC  & ASVspoof19 & 0.521 & 0.987  \\
 & ASVspoof19   & VSDC   & 0.57 & 0.99  \\
   \hline
 \multirow{2}{1.6cm}{MSRCC \cite{tapkir2018novel}} & VSDC  & ASVspoof19 & 0.49 & 0.99  \\
 & ASVspoof19   & VSDC   & 0.54 & 1.0  \\
  \hline
\multirow{2}{1.5cm}{RFCC \cite{saratxagainproceedings}}  & VSDC   &  ASVspoof19   & 0.44 &  0.99 \\
& ASVspoof19   & VSDC    & 0.67 & 1.0  \\
 \hline
 \multirow{2}{1.5cm}{IMFCC \cite{chakroborty2009improved}} & VSDC   & ASVspoof19   & 0.44 & 0.99  \\
 & ASVspoof19   & VSDC    & 0.61 &  1.0 \\
 \hline
\multirow{2}{1.5cm}{APGDF \cite{rajan2013using}} & VSDC   & ASVspoof19    & 0.34 & 0.82  \\
 & ASVspoof19   & VSDC     & 0.51 & 0.96 \\
 \hline
\multirow{2}{1.5cm}{CQCC \cite{Todisco2017ConstantQC}}  & VSDC   & ASVspoof19    & 0.50& 0.99  \\
& ASVspoof19   & VSDC     & 0.43 & 1.0  \\
 \hline
 \multirow{2}{1.5cm}{GTCC \cite{valero2012gammatone}} & VSDC  & ASVspoof19 & 0.49 & 0.99  \\
 & ASVspoof19   & VSDC   & 0.54 & 1.0  \\
  \hline
 \multirow{2}{1.5cm}{SSFC \cite{tapkir2018novel}} & VSDC  & ASVspoof19 & 0.51 & 0.99  \\
 & ASVspoof19   & VSDC   & 0.56 & 1.0  \\
  \hline
 \multirow{2}{1.5cm}{RPS \cite{saratxagainproceedings}} & VSDC  & ASVspoof19 & 0.50 & 0.96  \\
 & ASVspoof19   & VSDC   & 0.55 & 0.98  \\
  \hline
 \end{tabular} 
 \end{table}

\subsection{Overall performance of the SOTA countermeasures and a way forward}

One of the shortfalls in the current research into audio spoof detection is that cross-corpus evaluations are rarely done, limiting knowledge of the generalizability of the features and full countermeasure frameworks. To overcome this knowledge gap, this comparative analysis performs a cross-corpus evaluation, where one dataset is used for training and another for evaluation, to show the effectiveness of these features at generalized spoof detection across different conditions present in the respective datasets.

Based on the results of this study's experiments, we can infer that the no-countermeasures from the selected ones were consistent over a wide range of datasets and classifiers.
As shown in the preceding section, several countermeasures, such as CQCC \cite{Todisco2017ConstantQC} and RFCC \cite{saratxagainproceedings}, performed better overall in standalone testing of the countermeasures.
However, when the dataset and classifier were changed, the identical countermeasure failed to perform well. While PSRCC \cite{tapkir2018novel} and MSRCC \cite{tapkir2018novel} did not perform much better in standalone testing, they did significantly better in cross-corpus examination. After extensive testing, it is extremely difficult to select the best countermeasure from among those tested because some countermeasures performed well with one data set but failed with the other, while others performed in reverse with other classifiers. However, we may infer that classifiers and datasets have a significant influence on the present countermeasures. In addition, in standalone corpus testing with four alternative classifiers, the constant Q based CQCC \cite{Todisco2017ConstantQC} performed consistently with only a minor decline in performance across classifiers. The countermeasures PSRCC \cite{tapkir2018novel} and APGDF \cite{rajan2013using} scored marginally better than each other when the classifier varied and in cross-corpus evaluation. As a consequence of the extensive testing with various datasets and classifiers, the effectiveness of the countermeasures was called into question. It also demonstrates the significance of cross-corpus and variable classifier evaluation of proposed solutions.

\subsection{Performance Evaluation of recent deep learned and End-to-End solutions}
Following the observation of a performance drop in SOTA countermeasures, we conducted a cross-corpus sub-experiment utilizing contemporary deep learning, along with an end-to-end solution. In this subsection, the performance of recent deep learning and end-to-end solutions is evaluated. The results are in Table \ref{tab:cnn/deep learned models}. 
\par Several deep learning and end-to-end solutions have recently demonstrated outstanding performance in the recent ASV challenges \cite{jung2022aasist,rawboost, 9417604, aravind2020audio, li2021replay}. In the latest ASVspoof2021 competition, for instance, the end-to-end solution ASSIST \cite{jung2022aasist} achieved the best EER and min-TDCF on the ASVspoof2021-LA dataset. Similarly, \cite{9417604} and \cite{rawboost} also performed well in this task. Many participants, however, presented deep learning and end-to-end systems that had been trained and tested to mitigate only a single attack. To demonstrate the general effectiveness of the countermeasures, we performed an experiment in which we evaluated the higher performing solutions to the ASV challenges across corpora. Each solution trained and tested on ASVspoof2019-LA for the challenge, evaluated against the LA speech samples of the ASVspoof2021 corpus. Similarly, if the model's stated results were based on ASVspoof2021, we examined that model's performance against ASVspoof2019. The key reasoning behind this experiment was to evaluate and report the generalizability of the countermeasures. 
\par To measure generalizability, we selected the top-performing models with publicly available code. It was observed from the results that the top performer, ASSIST \cite{jung2022aasist}, and its companion model ASSIST-L, achieved EER values of 0.83\%  and 0.99\%, respectively, against the ASVspoof2019 dataset. However, the performance of these models deteriorated when tested against speech samples from the ASVspoof2021 dataset. Specifically, the EER scores increased to 15.84\% and 15.69\%, and the minimum t-DCFs rose from 0.027\% to 0.537\%, and from 0.030\% to 0.520\%, respectively. 
\par When tested against ASV spoof 2021, the EER for the solution proposed in \cite{9417604} increased from 2.17\% to 23.1\%, and the minimum TDCF increases from 0.055\% to 0.759\%, however the worst performance was observed in the case of the RawBoost solution proposed in \cite{rawboost}. Although the proposed solution performed optimally with the augmented data approach with EER score of 2.17\%, the EER and minimum t-DCF increased dramatically when tested on other datasets and without augmentation. In the case of ASVspoof2021 and without data augmentation, EER increased by 43\%, and the minimum t-DCF increased from 0.301\% to 0.903\%. 

\begin{table*}[t]
\caption{Computational details of the state-of-the-art countermeasures.}
\label{tab:cnn/deep learned models}
\centering
\begin{tabular}{*{7}{c}}
 \hline
  \multirow{2}{2cm}{Model} & \multicolumn{2}{c}{ASVspoof2019} & \multicolumn{2}{c}{ASVspoof2021} & \multicolumn{2}{c}{VSDC}  \\ 
                         & EER & min t-DCF & EER & min t-DCF & EER & min t-DCF \\
\hline
 ASSIST \cite{jung2022aasist} & 0.83 & 0.027 & 15.84 &  0.537 & 51.21 & --\\\hline
 ASSIST-L \cite{jung2022aasist} & 0.99  & 0.030 & 15.69 & 0.520 & 59.32 & --\\\hline
 RawBoost \cite{rawboost} & 2.17 & 0.055 & 23.1 & 0.759 & 64.90 & -- \\\hline
 One class Learning \cite{9417604} & 5.38 & 0.309 & 48.7 & 0.901  & 37.95 & -- \\\hline
 Res2Net34 \cite{li2021replay} & 1.67 & 0.058 & 48.00 & 0.994 & 29.00 & --\\\hline
 Res2Net50 \cite{li2021replay} & 1.63 & 0.049 & 39.00 & 0.999 &24.00  & --\\\hline
 ResNet34-LA \cite{aravind2020audio} & 10.53 & 0.19 & 26.55 & 0.34 & 40.23 & --\\\hline
 ResNet34-PA \cite{aravind2020audio} & 6.61 & 0.15 & 49.51 & 0.41 & 34.12 & --\\\hline

 \end{tabular}
\end{table*}

Although the adaptation of the Resnet-34 architecture published in \cite{aravind2020audio} reported an EER of 5.32\% against the ASVSpoof2019-LA dataset and 5.74\% against the ASVSpoof2019-PA dataset, we observed EERs of 10.53\% and 6.61\%, respectively, in our configured environment. When tested on ASVSpoof2021, performance dropped even further, with EER results of 26.55\% on LA and 49.51\% on PA. 
\par To further test the extensibility of this and other methods, we performed cross-attack, cross-corpus testing, training on ASVspoof2019-PA, and testing against ASVSpoof2021-DF, which achieved an EER 42.21\%. Similar performance deterioration was observed for \cite{aravind2020audio} on the distinct dataset; the EER rose to 48.00\% and 39.00\% for res2net34 and res2net50, respectively. The performance decreased when tested on the ASVspoof2021 dataset, and the worst performance was observed against the chained replay VSDC datasets. These scores imply that while these methods perform well on a specific dataset, the results do not transfer well to others. This test demonstrates the impact of formulating a solution based on a particular corpus, as the performance of deep learning and end-to-end solutions degrades significantly when presented with unknowns. Each static dataset differs and has distinct discrepancies due to varied factors, such as the ambient setting, gadgets, microphonic influence, and the spoofing technique used to produce spoof samples. Therefore, developing a spoofing countermeasure without taking into consideration the aforementioned factors may lead to performance degradation when tested across corpora. Thus, based on these results, we conclude that there is a dire need to concentrate on a broad approach that mitigates the aforementioned variances while developing robust voice spoofing defenses.

\subsection{Limitation and Future work}
Aside from the effectiveness of the presented work, there are certain limitations to this study. Due to contradictory technical issues and lack of code availability, only fourteen countermeasures were chosen to do the cross-corpus and multi-classifier evaluation. Moreover, in accordance with the advancement of the voice spoofing domain and deep learning methodologies such as CNNs, transformers, and so on, we compared four recent top-rated (by EER and min-TDCF scores on ASVspoof datasets) deep learning and end-to-end solutions with cross corpus evaluation. Despite the fact that just four models were examined, the findings motivate us to continue working in the deep learning and end-to-end arenas. Thus, we want to expand this work in the future to include the most recent deep learning/end-to-end solutions in order to provide an apples-to-apples comparison. Lastly, we intend to create a benchmark for comparing CNN and DNN-based architectures for the automatic speaker verification system.
\section{Existing Issues and Future Needs}
This section describes the challenges and issues observed during this review as well as the future requirements for developing dependable and secure anti-spoofing solutions.
\subsection{Need for Explainable AI in Audio Forensics}
Explainable AI (XAI) is an emerging field which emphasises the need for human understanding through what has traditionally been a black box process. An important gain over the interaction between humans and XAI is a level of trust in the process, which studies have shown does not currently exist \cite{needXAI}. Existing deepfake detection approaches are typically designed to perform batch analysis over a large dataset, however, when these techniques are employed in the field by journalists or law enforcement, there may only be a small set of videos available for analysis. A numerical score parallel to the probability of an audio or video being real or fake is not valuable to these users if it cannot be confirmed with an appropriate proof of the score. In those situations, it is very common to demand an explanation for the numerical score in order for the analysis to be believed before publication or utilization in a court of law, however, most deepfake detection methods lack such an explanation. If results for single interactions between humans and computers are to be accepted it is vitally important to involve XAI early in the process.

\subsection{Robust PAD for ASV in Smart Homes}
Because the Smart Home concept encourages hands-free automation, bio-metric technologies, and particularly automatic speech recognition, are the ideal way to regulate and personalize access. However, while contemporary spoofing systems are extremely descriptive and specific when detecting presentation attacks, current speaker identification systems may be vulnerable. Therefore, most of these PAD systems are ineffective in the current environment where the types of presentation attacks are unknown. When deployed to smart homes, the risk of attack increases significantly. Hence, there is an immediate need to build a strong PAD system to protect ASV systems in order to accelerate the safe use of ASVs in Smart Home applications. 

\subsection{What type of model is best for given scenario?}
Because countermeasures operate differently in varying conditions there is an urgent need to shift the research community's focus to environmental or scenario-based spoofing countermeasures. The environment is an essential part of spoofing detection countermeasures; for instance, several countermeasures may fail to provide adequate performance when surrounded by a loud environment or background noise. As a result, a thorough comparison study is required to find the best countermeasure for specific scenarios. 

\subsection{Need for a new Audio Forensics datasets}
Voice-controlled devices (VCDs) have transformed the Internet of Things (IoT), enabling the enhanced development, personalization, and  automation, of smart homes through voice activation. Unfortunately, these VCDs are currently largely unprotected, and can be exploited via spoofing attacks. Existing datasets, such as the Voice spoofing detection corpus (VSDC), ASVspoof 2017, ASVspoof 2019, and ReMASC, contain large-scale replay (i.e., single and multi-order replays) and basic cloning audio files, but newer, deep learning-based attacks may easily defeat current solutions, introducing a new level of complexity for anti-spoofing algorithms. Moreover, large-scale datasets like ASVspoof 2017 and ASVspoof 2019 lack cloning design parameters, which are an essential component of current VCDs. Therefore, an enhanced cloning dataset is needed in order to stay ahead of the curve of advanced voice cloning spoof attacks. For creation of such dataset, new generative algorithms such as diffusion models could be employed. Also, there is need to develop datasets that contains multiple forgeries at frame level in a single file. The new datasets also need to be unbiased with respect to gender and ethnicity. 
subsection{The role of Adversarial Machine Learning on existing Audio spoofing countermeasures}
High-performance spoofing countermeasure systems for automated speaker verification (ASV) have been presented during the ASVspoof sessions. However, the robustness of these systems in the face of adversarial attacks has yet to be thoroughly examined, while adversarial attacks pose a significant threat to ASV systems and their countermeasures. Currently, there are only a limited number of countermeasures in this domain. It is therefore necessary to develop a more adversarial-aware ASV system.

\subsection{Fairness-enabled voice spoofing countermeasures and ASV}
Fairness is defined as the lack of bias or preference toward a person or a group based on intrinsic or acquired attributes. Prior to their widespread commercialization, it is necessary to study the bias and impartiality of spoofing detection and speaker verification systems across populations. No quantitative evaluation has been reported to study the fairness (w.r.t gender, ethnicity etc) in the spoofing detectors across protected parameters, and even integrated voice countermeasure. To avoid the real-world consequences of a biased and flawed system that favors a single sub-group or class, it is vital to ensure the impartiality of data sets used for training and testing spoofing systems.

\subsection{Attack-aware Audio Forensics Countermeasures}
ASV (automated speaker verification) systems should always discard both non-target (voice produced by unrelated speakers) and spoofed (reprocessed or transformed) signals. However, little consideration has been given to how ASV systems should be changed when they encounter spoofing attacks, or even when they pair up with spoofing countermeasures, much less to how both systems might be jointly optimized. 

\subsection{Need for Cross-dataset Evaluation}
The need for cross dataset evaluation is important to consider when evaluating the performance of any aspect of an audio spoof detection system as shown in this paper. Cross dataset evaluation gives important insight into the robustness of the tested features and can provide additional information and direction toward a more general and complete ASV system.

\subsection{Physics of AI}
 Current deepfake detectors face challenges, particularly due to incomplete, sparse, and noisy data in training phases. There is a need to explore innovative AI architectures, algorithms, and approaches that “bake in” the physics, mathematics, and prior knowledge relevant to deep fakes and other forgeries. Embedding physics and prior knowledge, using knowledge-infused learning, into AI will help to overcome the challenges of sparse data and will facilitate the development of generative models that are causal and explanatory.

\section{Conclusion}
This research provides a comprehensive review of voice spoofing attacks, countermeasures proposed to counter certain attacks, publicly accessible datasets, and the criteria used to evaluate countermeasure performance in voice spoofing detection. This paper also includes an experimental comparison of the state-of-the-art (SOTA) voice spoofing countermeasures against multi-and cross-datasets and classifiers. Based on the results of the experiments, this study concludes that characteristics that better capture microphone signatures and harmonic distortions give improved detection performance for the identification of PA attacks. Furthermore, we show that the efficacy of the SOTA countermeasure varies significantly with changes in datasets and classifiers. Extensive testing reveals that these countermeasures, along with CNN and CNN-GRU classifiers, perform well. Detailed experiments also demonstrate the significance of cross-corpus evaluation for future voice spoofing technologies.  

\section{Reproducibility}
We have provided a GitHub repository\footnote{https://github.com/smileslab/Comparative-Analysis-Voice-Spoofing} that contains all the code and explanation needed to reproduce results of models evaluated to test their generalizability. The README.pdf file contains explanation of how to run the experiments.  

 \section{Acknowledgement}
 This material is based upon work supported by the National Science Foundation (NSF) under award number 1815724 and Michigan Transnationals Research and Commercialization (MTRAC), Advanced Computing Technologies (ACT) award number 292883. Any opinions, findings, and conclusions or recommendations expressed in this material are those of the author(s) and do not necessarily reflect the views of the NSF and MTRAC ACT.

\bibliographystyle{plain}
\bibliography{references.bib}
\end{document}